\documentclass[manuscript]{aastex}

\usepackage{natbib}
\usepackage{amssymb,amsmath}
\usepackage{graphicx}
\newcommand{\degree}{$^{\circ}$}

\newcommand{\kms}{km~s$^{-1}$}

\slugcomment{revised version for \apj}

\shorttitle{Arrival time calculation for circular ICME fronts}
\shortauthors{M\"ostl et al.}

\begin{document}




\title{Arrival time calculation for interplanetary coronal mass ejections with circular fronts and application to STEREO observations of the 2009 February 13 eruption}






\author{C. M\"ostl\altaffilmark{1,2}, T. Rollett\altaffilmark{1,2}, N. Lugaz\altaffilmark{3}, C. J. Farrugia\altaffilmark{4}, J. A. Davies\altaffilmark{5}, M. Temmer\altaffilmark{1,2}, A. M. Veronig\altaffilmark{2}, R. Harrison\altaffilmark{5}, S. Crothers\altaffilmark{5}, J. G. Luhmann\altaffilmark{6}, A. B. Galvin\altaffilmark{4}, T.~L. Zhang\altaffilmark{2}, W. Baumjohann\altaffilmark{2}, H. K. Biernat\altaffilmark{1,2}}

\affil{Institute of Physics, University of Graz, A-8010, Austria}
\affil{Space Research Institute, Austrian Academy of Sciences, Graz A-8042,
Austria}
\affil{Institute for Astronomy, University of Hawaii, HI 96822, USA}
\affil{Space Science Center and Dept. of Physics, University of New
Hampshire, Durham, NH 03824, USA }
\affil{RAL Space, Harwell Oxford, Didcot, OX11 0QX, UK.}
\affil{Space Sciences Laboratory, University of California, Berkeley, CA 94720, USA}
\email{christian.moestl@uni-graz.at}



\begin{abstract}


A goal of the NASA \emph{STEREO}  mission is to study the feasibility of forecasting the direction, arrival time and internal structure of solar coronal mass ejections (CMEs) from a vantage point outside the Sun-Earth line. Through a case study, we discuss the arrival time calculation of interplanetary CMEs (ICMEs) in the ecliptic plane using data from \emph{STEREO/SECCHI} at large elongations from the Sun in combination with different geometric assumptions about the ICME front shape (Fixed-$\Phi$ (FP): a point and Harmonic Mean (HM): a circle). 
 These forecasting techniques use single-spacecraft imaging data and are based on the assumptions of constant velocity and direction. We show that for the slow (350 km~s$^{-1}$) ICME on 2009 February 13--18, observed at quadrature by the two \emph{STEREO} spacecraft, the results for the arrival time given by the HM approximation are more accurate by 12 hours than those for FP in comparison to in situ observations of solar wind plasma and magnetic field parameters by \emph{STEREO/IMPACT/PLASTIC}, and by 6 hours for the arrival time at Venus Express (MAG). We propose that the improvement is directly related to the ICME front shape being more accurately described by HM for an ICME with a low inclination of its symmetry axis to the ecliptic.  In this case the ICME has to be tracked to $>30^{\circ}$ elongation to get arrival time errors $< \pm$ 5 hours. A newly derived formula for calculating arrival times with the HM method is also useful for a triangulation technique assuming the same geometry.
\end{abstract}





\section{Introduction}

Coronal mass ejections (CMEs) are episodic expulsions of huge amounts of plasma and magnetic fields from the solar corona \citep[e.g.][]{how85}. CMEs form a centerpiece of space weather research since they are the source of the strongest disturbances in the Earth's magnetosphere \citep[e.g.][]{tsu88,far93,gos93,bot06,zha07}. Besides other detrimental effects, the resulting geomagnetic storms can knock out communication satellites, lead to errors in global positioning systems and induce currents on the ground leading to wide-spread electricity blackouts \citep[e.g.\ report by the][]{nat08}. The NASA \emph{STEREO} (Solar TErestrial RElations Observatory) mission \citep{kai08} was launched in October 2006 and is designed to enhance our understanding of the initiation and propagation of CMEs, and thus foster our ability to forecast them, using simultaneous imaging in EUV and white-light coronagraphs \citep[][]{how08}, and sampling the solar wind magnetic fields and plasma using the in situ instruments PLASTIC \citep{gal08} and IMPACT \citep{luh08}. Additional to the novel three viewpoints through in situ instruments and coronagraphs (including SOHO / LASCO and the Wind/ACE in situ measurements near Earth), both \emph{STEREO} spacecraft are equipped with Heliospheric imagers \citep[HIs,][]{eyl09} which can directly map the complete evolution of  Earth-directed transients (such as ICMEs) in the inner heliosphere ($< 1$~AU) from a vantage point away from the Sun-Earth line. By definition, we use the term ICME for the structure after it has left any coronagraph's field of view (FoV), i.e.\ we use it for both the in situ and HI observations.

Especially useful is the following concept based on earlier work by \cite{she99} and demonstrated with STEREO/HI and SMEI by e.g.\ \cite{rou08,rou09},  \cite{dav09}, \cite{moe09c,moe10} and \cite{how09b}: if an ICME propagates with constant velocity and direction to large angles with respect to the Sun, an observer perceives deceptive angular accelerations/decelerations depending on its propagation direction relative to the observer. This effect can be used to assess the direction, velocity and arrival time of an ICME by fitting known theoretical profiles to real-time \citep{dav11} or level 2 HI or SMEI data of the ICME's elongation.  There exist two practical realizations of this concept: (1) the original technique \citep{rou08} which assumes a point-like or negligibly narrow shape of the CME front (also called the Fixed-$\Phi$ approximation; we call this Fixed-$\Phi$ fitting or FPF), or the method by \cite{lug10b} which assumes a circular CME front, attached to the Sun at all times at one end (Harmonic Mean Fitting or HMF). It has to be pointed out that there also exist more complex approaches to the same problem like the T-H model \citep{how09a,how09b,tap09} where a large set of pre-existing solutions is fitted to the observations, also useful for real time application, and the white-light rendering method by \cite{woo09}. In this study we use the FPF and HMF methods, and we denote the formulas for converting elongation to distance with FP/HM and the inverted method useful for predicting ICME parameters  with FPF/HMF. 

The big advantage of the fitting techniques is that they are simple and fast to apply in real time, and they can be used for single spacecraft HI observations, which will most likely be available on future solar missions or an observatory at the L5 point in the Sun-Earth system, 60$^{\circ}$ behind the Earth  \citep{sch96,gop11}. The big disadvantage is that these methods assume constant speed $V$ and direction $\phi$. In general, fast and more geoeffective ICMEs are expected to decelerate because the solar wind they run into is usually slower and exerts a drag force, and ICME directions can be influenced by high speed solar wind streams \citep[e.g.][]{gop09}. Thus, for a real-time application, the resulting speeds and directions will only be averages of the part where the elongation profiles are measured, and their errors have to be carefully assessed. Using two HI observatories, the assumptions of constant velocity and direction can be dropped \citep{liu10,liu10b, lug10}, and again there exist two limiting cases for the ``extreme'' geometric assumptions of a point or a circle: triangulation \cite[TR;][]{liu10} and tangent-to-a-sphere \citep[TAS;][]{lug10}. An excellent summary of these techniques is given by \cite{liu10b}.

However, what is missing so far is a way how to calculate the speeds and arrival times for the methods assuming a circular front shape (HMF and TAS). It is the aim of this paper to derive an analytical formula which gives the arrival time and velocity of an ICME using the HMF and TAS methods. The best way to independently assess the validity of any of these techniques is to check the results against in situ magnetic field and plasma measurements of ICMEs. Thus, additionally, we assess the validity of our new formula by comparing the results of the FPF and HMF methods for an ICME event on 2009 February 13--18 \citep[see also][on the EUV wave]{pat09,kie09,coh09}, which we believe to be extremely well suited for this purpose for two reasons: the CME had a (1) practically constant velocity all the way in the HI FoVs and (2) it was propagating into a ``simple'' slow background solar wind, so that any differences in the results by the FPF / HMF methods should be based only on the assumption of the different geometry rather than other effects.

The paper is structured as follows: in Section 2, we present an alternative version of the \cite{lug10b} formula for inverting the Harmonic Mean approximation for the global ICME leading edge shape, derived in Appendix~A. Also, an associated correction formula for calculating the arrival time with HMF is discussed in Section 2 and derived in Appendix~B. In Section 3, we discuss  \emph{STEREO-A}/HI observations of the 2009 February 13--18 ICME event and application of the HMF and FPF methods. We check these results against the ICME in situ observations by \emph{STEREO-B} and Venus Express in Section 4. We summarize our conclusions in Section 5.

\section{Methods for fitting the time-elongation profiles of ICMEs}

\subsection{Fixed-$\Phi$ and Harmonic Mean}
For completeness, we revisit in this section methods to calculate a constant ICME direction and velocity from the observed elongation variation with time $\epsilon(t)$ of a feature (often the ICME leading edge) in the HI field of views. For two excellent in-depth summaries see \cite{lug10b} and \cite{liu10b}. We will restrict ourselves here to a short discussion and introduce definitions to be used in the following case study. We present in this section (1) a new alternative version of a formula which can be used for fitting time-elongation tracks of transients assuming a circular front shape (Harmonic Mean Fitting or HMF, \cite{lug10b}), and (2) a new geometrically consistent formula for deriving the ICME arrival time at a given location in the heliosphere with  HMF. Both formulas are derived in the Appendix to this paper.
 
\begin{figure}
\epsscale{.80}
\plotone{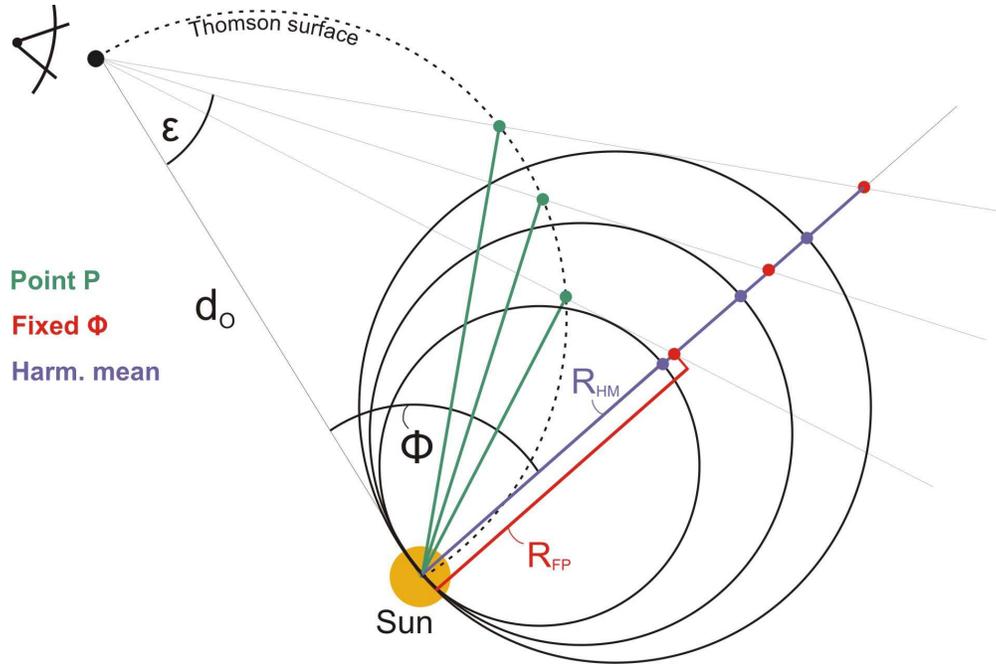}
\caption{Illustration of the different elongation-to-distance conversion methods. 
From the top left, an observer looks towards a circular ICME front. The green (PP), red (FP) and blue (HM) dots show where the observer would
calculate the radial distance of the ICME from the Sun using the indicated methods. In this case, the observer used the same direction $\phi$ for the FP and HM conversion equations, which leads to $R_{FP} > R_{HM}$ for the same elongation value. }\label{fig:methods_sketch}
\end{figure}

The effects of observing features of ICMEs at large angles to the Sun were introduced by \cite{she99}. Basically, the assumption that the CMEs intensity signal originates from the plane-of-sky, as valid in the coronagraph field of views close to the Sun,  is not valid for larger angles from the Sun \citep{vou06,how09a}. The radial distance $R_{FP}$ from the Sun of a point-like transient moving at a longitude $\phi$ ($\phi > 0$ is defined as westward) relative to the observer, who is situated at a distance $d_o$ from the Sun, which is observed at time $t$ at an elongation angle $\epsilon$, is given by \citeauthor{she99}~(\citeyear[][their equation B1]{she99}) as

\begin{equation}\label{equ:fpconversion_vorn}
    R_{FP}(t)= \frac{d_o\sin \epsilon(t)}{\sin(\epsilon(t)+\phi)}.   \\
\end{equation}
This is also known as the  Fixed-$\phi$ (FP) approximation \citep{kah07}. These definitions are visualized in Figure~\ref{fig:methods_sketch}. Solving for $\epsilon(t)$ leads to
\begin{equation}\label{equ:fpfit_vorn}
    \epsilon (t) = \arctan \left ( \frac{V_{FP} t \sin(\phi_{FP})}{d - V_{FP} t \cos (\phi_{FP})} \right),
\end{equation}
first written down explicitly by \cite{rou08}. Here we have already replaced the distance $R_{FP}$ by $V_{FP} t$. This formula is thus only based on a constant velocity $V_{FP}$ and a constant direction $\phi_{FP}$. For comparison with real data, it is also necessary to obtain a launch time $t_{0FP}$ on the Sun, for which $\epsilon(t_{0FP})=0$\footnote{We note that for a consistent comparison of the launch time to solar surface signatures any $\epsilon(t_{0})$ should be set to the angular size of the Sun's radius ($\approx 0.26^{\circ}$ for an observer at 1 AU), but to simplify the arrival time calculation $\epsilon(t_{0})=0$ in this paper.}. Equation (\ref{equ:fpfit_vorn}) has the obvious advantage that a theoretical elongation profile can be fitted in a least-squares sense to the observed values of $\epsilon(t)$, which results in the 3 parameters $V_{FP}, \phi_{FP}$, and $t_{0FP}$ \citep[e.g.][]{rou08,rou09,dav09,dav10,moe09c,moe10,lug10b}. We further call this the Fixed-$\phi$-Fitting or FPF approach. We implemented the fitting by using a downhill simplex method (\cite{nel65}, IDL function \texttt{amoeba.pro}) which is, from our experience, robust for this three-parameter minimization. It can also be done by creating an ``error map'' for $V_{FP}$ and $\phi_{FP}$ and looking for the minimum residual between the theoretical and observed ICME track \citep{wil09,lug10b}.

The same procedure can be applied using the so-called Harmonic Mean approach. \cite{lug09a} derived a formula to convert elongation to distance, similar to Equation~(\ref{equ:fpfit_vorn}), but under the different assumption that the CME front can be approximated by a circle, which is attached to the Sun at all times at one end. This definition is visualized in Figure~1.  It is further assumed that the observer always measures the resulting elongation angle by looking along the tangent to this circle \citep[see Figure 1 and][]{lug09a,how09a}. The radial distance of the circle's apex $R_{HM}$, the point with largest distance from the Sun, is given by \cite{lug09a}:

\begin{equation}\label{equ:hmconversion}
    R_{HM}(t)=  \frac{2 d_o \sin \epsilon(t)}{1+\sin(\epsilon(t)+\phi)}.   \\
\end{equation}
An equivalent formula was also derived by \citeauthor{how09a}~(\citeyear[their equation 41]{how09a}).  We derived a new version of the inversion of this formula in Appendix A, compared to the one given by \cite{lug10b}. Setting $R_{HM}=V_{HM}t$ and defining
\begin{equation}
   a=\frac{2d_o}{V_{HM} t}-\cos{\phi_{HM}}; \quad   b=\sin(\phi_{HM}), \\
\end{equation}
leads to
\begin{equation}\label{equ:hmfit}
   \epsilon (t) = \arccos ( \frac{-b + a \sqrt{a^2+b^2-1}}{a^2+b^2}).  \\
\end{equation}

This has only one term in contrast to the formula with two terms given by \cite{lug10b}. The reason we use this formula instead of Equation~(1) by \cite{lug10b} is that we found it works more robust when running the downhill simplex method for the fitting, which we use in contrast to \cite{lug10b}, who constructed error maps for a grid of $V_{HM}$ and $\phi_{HM}$ values. It is our purpose to find out if the two approaches may produce different results in future studies and to further check the robustness of our fitting method, but in this paper we restrict ourselves to discussing the results of the simplex method alone. 

\subsection{Arrival time calculation using fitting methods}

In this section we summarize the calculation of the arrival time $t_a$ of an ICME at a particular planet or spacecraft for both fitting methods.  Beside the constant direction, the FPF and HMF methods result in a constant velocity $V_{FP}$ / $V_{HM}$ and a launch time $t_{0FP}$ / $t_{0HM}$ on the solar surface, for which $\epsilon(t_0)=0$.
For the FPF method the arrival time is  then simply given by

\begin{equation}\label{equ:arrtime_vorn}
 t_{aFP}=t_{0FP} + d_i/V_{FP},
\end{equation}
with $d_i$ being the radial distance of the in situ observing spacecraft from the Sun.
For the FPF method, it is assumed that the ICME's white-light emission arises from a single plasma element, or in reality that the ICME front is of negligible small extent along the line-of-sight. But what does a ``small extent" exactly mean? In this respect it is necessary to define for which  cases an ICME with a direction $\phi_{FP}$ is actually expected to hit a spacecraft. It has been assumed in previous studies \citep[e.g.][]{dav09,dav10} that if the direction $\phi_{FP}$ matches the in situ spacecraft longitude to within a few degrees, the ICME is expected to hit the spacecraft and the arrival time can be calculated by using Equation (\ref{equ:arrtime_vorn}). This actually implies that the front shape of the ICME for the FPF method is a small segment (extending a few degrees in longitude) of a circle with radial distance of 1~AU from the Sun around the direction $\phi_{FP}$. But for larger angular separations between the in situ spacecraft and the ICME direction this definition runs into troubles because it is unclear if the ICME is expected to hit the spacecraft at all. Clearly, only a statistical study using many ICMEs will determine where to set the boundary where a prediction should be issued that an ICME will hit a particular planet or spacecraft at a certain longitudinal separation angle (which we further call $\Delta$) between the ICME direction and the target position.

This problem is less evident for the HMF method because the ICME front shape is  well defined as an expanding circle attached to the Sun at one end. However, the opposite problem arises here compared to the FPF method: there it is not clear if a spacecraft is hit at all by an ICME, but here sooner or later \emph{every }in situ spacecraft should detect the ICME front which is positioned at a longitude of $\phi_{HM} \pm 90^\circ$, i.e., in the half space centered on $\phi_{HM}$. Nevertheless, the well-defined circular geometry directly leads to an analytical arrival time calculation, which we derive in Appendix~B. To calculate an ICME arrival time at a given heliospheric position which is geometrically consistent with the circular front,  it is necessary to correct for the fact that all parts of the circular front except for the apex will always arrive later at a given radial distance from the Sun than the apex (see also Figure~\ref{fig:correction}a). To calculate the arrival time of an ICME at a given heliospheric distance $d_i$ and  at a longitudinal separation $\Delta$ with the HMF method, an arrival time correction $t_c$ must be added to the arrival time of the ICME apex:
\begin{equation}\label{equ:arrtimehm1}
 t_{aHM}=t_{0HM} + \frac{d_i}{V_{HM}}+t_c.
\end{equation}
As derived in Appendix B, this can be reduced to the simple  equation
\begin{equation}\label{equ:arrtimehm_simple_vorn}
 t_{aHM}=t_{0HM} + \frac{d_i}{ V_{HM}\cos(\Delta)}.
\end{equation}
It is seen in a straightforward way that the circular HM geometry leads to the elementary result that the speed of the ICME flank at a given angle $\Delta$ to the ICME apex is just reduced by a factor $\cos(\Delta)$, for this particular geometry. This is especially important for comparison to in situ measurements of the solar wind bulk speed when using the HMF method, where the speed $V_{HMI}=V_{HM} \cos(\Delta)$ has to be used, in contrast to FPF, where just $V_{FP}$ has to be taken.

 Additionally, there exists a triangulation version of the HMF approach, which was called the tangent-to-a-sphere (TAS) method \cite[Model 1 in][]{lug10}.  In contrast to the single-spacecraft fitting methods, using two $\epsilon(t)$ functions gained by widely separated observers makes it possible to relax the assumptions of constant  speed $V$ and direction $\phi$ \citep[see also][]{liu10,liu10b}. For calculating the ICME arrival time with TAS, Equation (\ref{equ:arrtimehm_simple_vorn}) also has to be used, but because $V_{TAS}(t)$ and $\phi_{TAS}(t)$ are now functions of time a final speed and final direction have to be put into Equation (\ref{equ:arrtimehm_simple_vorn}), i.e.\ the resulting speed and direction values at the last time step both $\epsilon(t)$ functions could be determined. We imagine that the final direction $\phi_{TAS}$ is simply the result of the last $\epsilon(t)$ pair, but for the terminal propagation speed $V_{TAS}$ averages of the last few hours need to be taken because of the fluctuations resulting from the derivatives \citep[see e.g.\ Figure 4 in][]{liu10}. Alternatively,  the arrival time correction $t_c$, which is given by Equation (\ref{equ:arrtimehm_niceunits}) in the Appendix B in easy-to-use units, needs to be added to an already calculated arrival time of the ICME apex at a given radial distance from the Sun.

Before proceeding to a data example, we also note that we extract the ICME track from \emph{STEREO-A}/HI images in the same plane (the ecliptic) as the main in situ observing spacecraft (\emph{STEREO-B}) is situated. In this way we avoid any effects based on observations and positions yielding from different latitudes.


\section{Case study of the 2009 February 13--18 ICME}


On 2009 February 13 a CME erupted from a small bipolar active region (AR 11012) at position S04E46 as seen from Earth \citep[][Miklenic et al. 2011, submitted to Solar Physics]{kie09}. It was associated with a GOES class B2.3 flare peaking at 05:46 UT and an EUV wave, the latter being discussed by \cite{kie09,pat09} and \cite{coh09}. Between 2009 Feb 13--18 \emph{STEREO-B} was situated on average at 1.0033~AU and at a longitude of $-47.6$~degrees east of the Earth in Heliocentric Earth Ecliptic (HEE) coordinates, \emph{STEREO-A} at 43.5 degrees west of the Earth at 0.9643~AU, and their separation in HEE latitude was negligible ($<0.3^{\circ}$). This means that the event was seen at the limb from \emph{STEREO-A} and almost exactly at disk center by \emph{STEREO-B}, making this  a practically  perfect quadrature configuration, as the \emph{STEREO} longitudinal separation was 91 degrees. The two probes are  equipped with instruments \citep[SECCHI,][]{how08} to image the Sun's corona in the extreme ultraviolet spectrum (EUVI), and the inner (COR1, 1.5--4~R$_\odot$) and outer corona (COR2, 2.5--15~R$_\odot$) in white-light. Images of the heliosphere, also in white-light, are obtained by the Heliospheric Imagers \citep[SECCHI/HI; ][]{eyl09}, with HI1 covering 4--24 degree elongation from the Sun and HI2 18.7--88.7 degrees. These values are valid for the ecliptic plane, at which the optical axes of the HI telescopes are pointing nominally. STEREO is also able to probe the solar wind at the position of each spacecraft itself with two in situ instruments: IMPACT \citep{luh08} and PLASTIC \citep{gal08}.

Miklenic et al.\ (2011, submitted to Solar Physics) discussed the early acceleration phase of the event and found that the CME reached its propagation velocity of $350$~km~s$^{-1}$ already at around $2.5$~R$_{\odot}$, which is still in the field-of-view (FoV) of the STEREO/COR1A coronagraph and thus rather close to the Sun and well below the HI FoVs. The ICME was detected in situ by \emph{STEREO-B} and Venus Express (see Section 3.2), with an observed average proton bulk speed at \emph{STEREO-B}  also of $\approx 350$~km~s$^{-1}$. The state of the background solar wind for this event is discussed by Temmer et al.\ (2011, submitted to ApJ) using the WSA/MAS+ENLIL solar wind model \citep{ods03}, who found a rather simple state of the heliosphere around the time and location of the eruption, with a small high-speed-stream to the east of the source region. \emph{STEREO-B} EUVI images also show that the AR 11012 was the only one on the solar disk viewed from the location of \emph{STEREO-B}.

In summary, the CME was slow, reached its terminal velocity very close to the Sun and propagated into a simple, slow solar wind, approximately maintaining its constant velocity from 2.5 R$_{\odot}$ up to 1~AU (215 R$_{\odot}$) (see also Rollett et al., 2011, submitted to Solar Physics). These characteristics make the event especially well suitable for a comparison between the HMF and FPF methods, because the ICMEs propagation in the HI FoVs, where the methods are applied, will not be affected strongly by any interaction with fast streams from coronal holes \citep[e.g.][]{gop09} or other ICMEs, thus minimizing the effects of a possible deflection or acceleration during its propagation through the \emph{STEREO-A} FoVs, which would be in conflict with the FPF/HMF assumptions of constant velocity and constant direction. Thus any differences  in the results may be safely ascribed to the strongly different assumptions on the shape of the ICME front as used for FPF/HMF.

\subsection{STEREO-A heliospheric imaging observations}

\begin{figure}
\epsscale{0.85}
\plotone{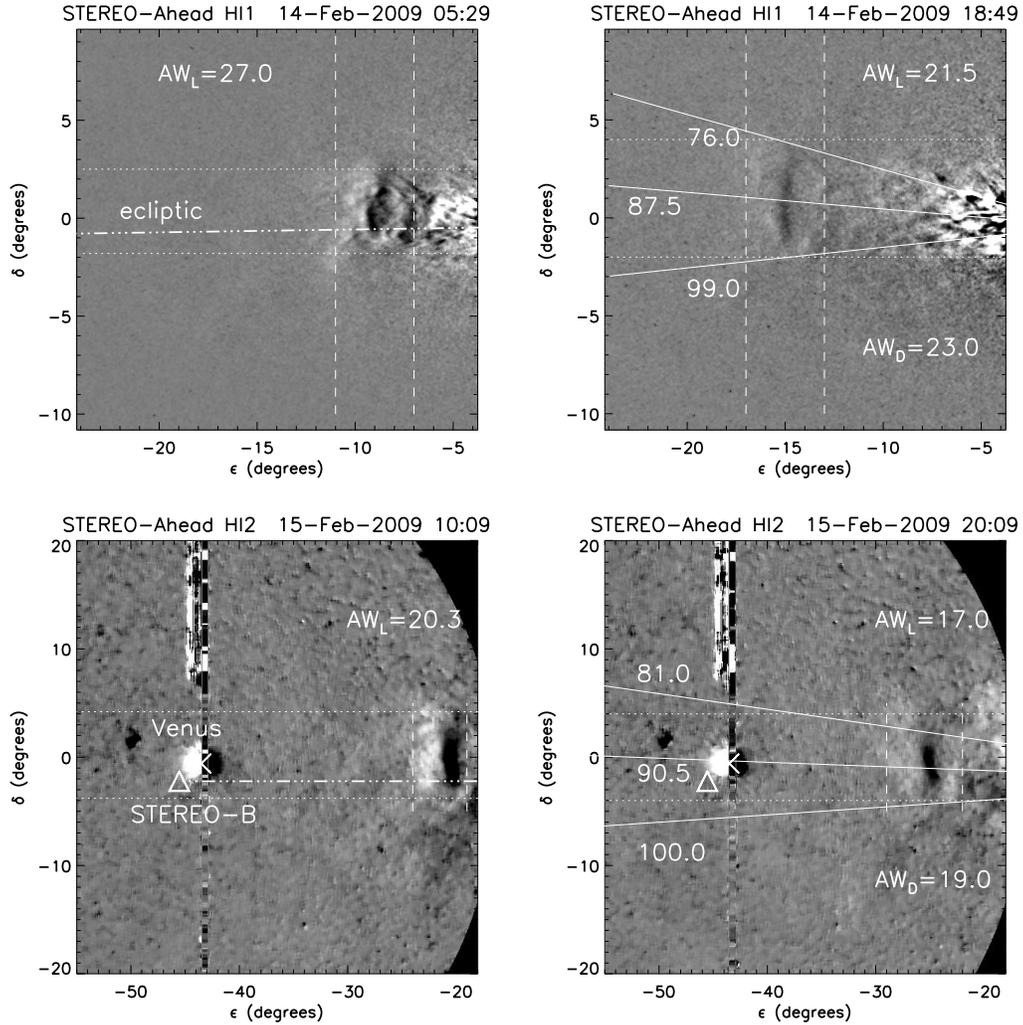}
\caption{Evolution of the ICME in \emph{STEREO-A} HI1 (top 2 images) and HI2
 (bottom 2 images). The ICME is delimited in elongation $\epsilon$ and elevation $\delta$ by dashed and dotted lines, respectively. These coordinates represent longitude ($\epsilon$) and latitude ($\delta$) in a Helioprojective Cartesian (HPC) coordinate system. In the left two images the ecliptic (in the plane-of-sky of \emph{STEREO-A}) is plotted as a dashed-dot-dot line. The positions of \emph{STEREO-B} (triangle) and Venus (cross) are indicated in the bottom two images. In the right two images the upper, central, and lower position angles of the ICME are indicated, again for the plane-of-sky of \emph{STEREO-A}, giving a directly measured angular width of $AW_D=23^{\circ}$ (HI1) and $AW_D=19^{\circ}$ (HI2). The angular widths $AW_L$ in all images are based on the formula by \cite{lyn10}, and are quite similar to the $AW_D$ values. The small angular width and the morphology are consistent with a CME viewed edge-on, so its axis of symmetry points out of the paper plane. This figure is also available as an mpeg animation in the electronic edition of the {\it Astrophysical Journal}.}\label{fig:hi_evolution}
\end{figure}

In Figure~\ref{fig:hi_evolution} the top 2 images show running differences of the ICMEs evolution in \emph{STEREO}-HI1A and the bottom 2 images show HI2A. The intensities in the images are related to Thomson-scattering of white-light solar radiation off solar wind electrons, and the incident intensity for a given observer maximizes  along the line-of-sight at the Thomson-surface \citep[][see also Figure \ref{fig:3d}]{vou06,how09a}. We encourage the reader to also look at the HI1A/HI2A movie accompanying the paper in the electronic edition of \emph{The Astrophysical Journal}. Overall, the CME is rather small and faint. It is actually so faint that it cannot be discerned in \emph{STEREO-B} HI movies, thus the observations limit our study to using single-spacecraft HI methods, so it is not possible to use any triangulation approaches. In the first image, it is seen that the ICME clearly has a ringlike shape with a small angular width of roughly $20^{\circ}$, which is consistent with a view of a croissant-shaped ICME edge-on \citep{the09,woo09}, i.e.\ we look along the axis of the ICME, which is pointing out of the paper plane.  The second density enhancement at its sunward end disappears around $-15^{\circ}$ elongation, leaving only the leading edge visible.

We estimated the ICMEs angular width ($AW$) in HI using the method of \cite{lyn10} (results denoted by $AW_L$) and additionally by a direct measurement in the plane-of-sky (POS) of \emph{STEREO-A} (denoted by  $AW_D$). The results are indicated in Figure~\ref{fig:hi_evolution} for each frame and show that between early Feb 14 and late Feb 15 $AW$ stayed between 19 and 27 degrees, which is practically constant. In the accompanying online movie it is also seen that neither the angular width nor the latitudinal propagation direction and the morphology significantly change during the full propagation in the HI FoVs. Later than early Feb 16 the ICME becomes very faint in a running difference image, but its leading edge still can be seen in the Jmap (see below and Figure~\ref{fig:jplot_comparefit}b).


In the right two images of Figure~\ref{fig:hi_evolution} it is also seen that the ICME propagates along a central position angle (PA) about 2 degrees north of the ecliptic plane, and the latter is drawn at a PA of 92.0$^{\circ}$ in the  \emph{STEREO-A} POS. The ICMEs central PA in the HI1 frame is 87.5\degree, and 90.5$^{\circ}$  in the HI2 frame. Thus, in latitude, the ICME propagates outward close to the ecliptic, and thus close to \emph{STEREO-B} (at PA $ =92.0$\degree) and Venus (at PA $ =90.3$\degree). The heliospheric longitude of \emph{STEREO-B} is only 1$^{\circ}$ to the east of the Solar source region of the CME, and Venus is located 21$^{\circ}$ to the west of it, at a radial distance of 0.7185~AU and a HEE longitude of $-25.4$~degrees (roughly halfway between \emph{STEREO-B}  and the Earth).  Thus, judging from the location of the Solar source region, the spacecraft positions and the symmetry axis close to the ecliptic plane the ICME is expected to hit both \emph{STEREO-B}  and Venus. 

Figure~\ref{fig:jplot_comparefit}a shows the proton density $N_p$ observed by \emph{STEREO-B}.
Panel (b) shows a Jmap extracted along the ecliptic plane. This is done by stacking vertically the median of the central rows $\pm 16$ pixels in the \emph{STEREO-A} HI1/2 running difference images \citep[for further details see][]{dav09b}.  It is seen that on early 2009 February 18 a rise in $N_p$ corresponds to the arrival of a track in the Jmap which is the leading edge of the ICME seen in Figure~\ref{fig:hi_evolution}. The track in the Jmap was manually selected by choosing 30 points along the front of the leading edge, for 10 times. From the uncertainties associated with the manual selection of the points we will discuss how the results of the fitting procedures are affected, so all systematic errors are assumed to arise from the manual selection only. The measured points were then interpolated onto an equidistant temporal grid consisting also consisting of 30 points (see Figure~\ref{fig:jplot_comparefit}c).

\subsubsection{Results of the fitting methods}

\begin{figure}
\epsscale{.95}
\plotone{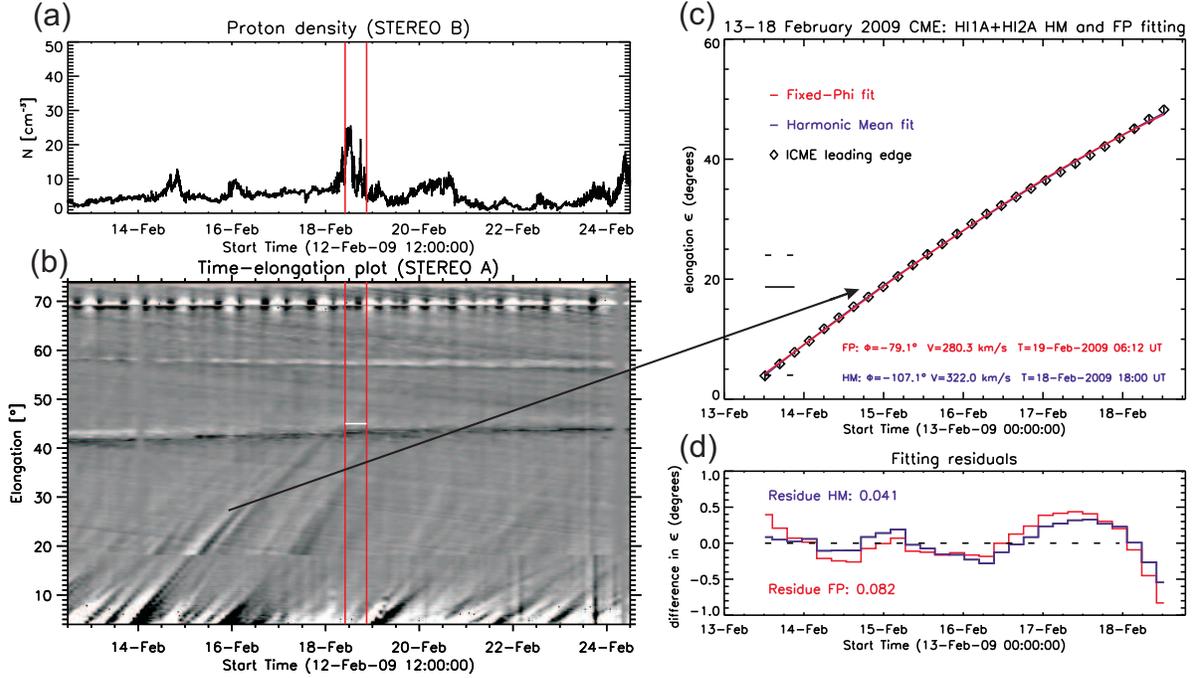}
\caption{(a) Proton density observed by \emph{STEREO-B/PLASTIC}. Enhanced densities (up to 25 protons cm$^{-3}$) in front of the flux rope are delimited by two vertical lines, with the left one setting the arrival time.  (b) Jmap along the ecliptic plane with the elongation of \emph{STEREO-B} marked as a white horizontal line. It is seen that a track, starting on Feb 13, reaches the elongation of \emph{STEREO-B} and Venus on early Feb 18, which matches very well the single density jump observed by \emph{STEREO-B}. (c) Best HMF / FPF fits for the ICME leading edge. Both functions are able to reproduce the observed function $\epsilon(t)$. (d) Residuals between the fitting functions and the observations. The time development of $\epsilon(t_i)-\epsilon_{fit}(t_i)$ is similar for both techniques and the summarized residue $\sigma^2 $ for HMF is half the value for FPF.}\label{fig:jplot_comparefit}
\end{figure}

We now applied the FPF and HMF methods, as introduced in Section~2, to the measured ICME track. Figure~\ref{fig:jplot_comparefit}c shows the results. In panel (c), the fitting functions (red: FPF; blue: HMF) are shown along with the observed $\epsilon(t)$ and the error bars arising from the manual selection of the track. Both methods reproduce $\epsilon(t)$ very well, but the results for the free parameters  differ considerably (see also Table 1):
\begin{align}
\phi_{FP}=-79^{\circ}, \quad   V_{FP}=280  \text{ km s}^{-1}, \\
\phi_{HM}=-107^{\circ}, \quad  V_{HM}=322 \text{ km s}^{-1},
\end{align}
 so the HMF direction is $28$ degrees eastward (or further away from the observer) compared to FPF. This has been noted by \cite{lug10b} to be a general feature -- especially for a limb event the observer will mostly see the flank of the ICME in HI because of the scattering at the Thomson surface. By ``flank'' we denote those parts of the ICME front which are closer to the Sun compared to the apex, the point along the circle with the greatest distance from the Sun. The HMF method corrects this effect and will put the ICME apex direction always farther away from the observer compared to FPF \citep{lug10b}. In this case the directions for FPF is 12~degrees west of the longitude of \emph{STEREO-B} and HMF 16~degrees east.

The expected in situ speed at the location of \emph{STEREO-B} of  $V_{HMI}= 310$~km~s$^{-1}$ is 30~km~s$^{-1}$ faster compared to FPF. We attribute this again to the fact that for a limb ICME event, especially one with low axis inclination to the ecliptic (to be justified further in the next section), more and more of the ICME flank is sampled. For a circular front, points at the flank will always move slower than the ICME apex, and again the HMF technique corrects this effect which results in a higher speed compared to FPF \citep{lug11}.  Because the launch times $t_0$ are quite similar (see Table 1 for a summary), the different speeds lead to quite different arrival times: due to its higher speed, $t_{aHM}$, which already includes the arrival time correction we presented in Section~2, is about 12 hours earlier compared to FPF, which is a considerable difference.

 In Figure~\ref{fig:jplot_comparefit}d we demonstrate how well both fits are able to reproduce the observed function $\epsilon(t)$. To this end, we use the fitting residual \citep{wil09,lug10b}
\begin{equation}\label{eq:fitresiduals}
    \sigma^2= 1/N \sum_{i=1}^{N} (\epsilon(t_i)-\epsilon_{fit}(t_i))^2
\end{equation}
with $N$ being the number of data points along the ICME track and $\epsilon_{fit}$ the calculated elongation value for the time step $t_i$ using the given set of fitting results $V$, $\phi$ and $t_{0}$. For the least-squares minimization using the downhill-simplex-method, a slightly modified $\sigma^2$ is used, where we replaced both $\epsilon$ terms with their norms $|\epsilon(t_i)|$ and $|\epsilon_{fit}(t_i)|$, which lead to a more robust minimization than Equation~(\ref{eq:fitresiduals}). From Figure~\ref{fig:jplot_comparefit}d, where we plotted $\epsilon(t_i)-\epsilon_{fit}(t_i)$ for every $\epsilon(t)$ measurement, it is seen that both functions are able to almost equally well describe the observed elongation variation of the ICME leading edge. The residual $\sigma^2$ is a factor of 2 higher for FPF than for HMF, and both residual functions follow each other very closely, with the strongest deviations to the observed elongations at the end of the track. Nevertheless, based on the fits one cannot clearly distinguish between the two methods based on the consideration that a goodness-of-fit is better for one method as compared to the other. A real-time space weather predictor would thus not be able to distinguish if one of the predictions should be preferred over the other because he/she is able to fit the $\epsilon(t)$ function with almost the same accuracy with both methods. To the contrary, both functions have three free parameters and are flexible enough to fit $\epsilon(t)$ almost perfectly, but the results of the inversion differ quite greatly due to the vastly different assumed geometries. Thus, the only way to distinguish between the success of the two methods is to look if the independent in situ observations, available for this event even by two spacecraft, are consistent with the results derived from the heliospheric images. But before we proceed to discuss the in situ data, we assess the errors involved in the derived parameters.

\subsubsection{Error analysis}\label{errorsection}

How do the errors arising from the manual selection of the ICME track in the Jmap impact the results of the FPF and HMF methods? To this end, we constructed two additional tracks
\begin{gather}\label{equ:errors}
    \epsilon_+(t) = \epsilon(t) + 2\sigma_{\epsilon}(t) \\
    \epsilon_-(t) = \epsilon(t) - 2\sigma_{\epsilon}(t)
 \end{gather}
with $\sigma_{\epsilon}(t)$ being the standard deviation of $\epsilon(t)$. Thus  $\epsilon_+(t)$ / $\epsilon_-(t)$ is the track with the maximum positive / negative deviation from $\epsilon(t)$. We again fitted the two tracks of maximum deviation with the FPF and HMF methods and averaged the difference between these results and the ones for $\epsilon(t)$ (Table~1) to obtain an error for $V$, $\phi$, $t_0$ and $t_a$. For the latter the HMF arrival time correction has to be included. We found a mean standard deviation of $2\sigma_{\epsilon}=0.51^{\circ}$ arising from the manual selection of points, thus the systematic error associated with the manual measurement in the Jmap is about $\pm 0.5^{\circ}$. This error affected the direction as $\phi_{FP}= -82^{\circ} \pm 1.8$ and $\phi_{HM}= -111^{\circ} \pm 2.8$. The errors in the speeds were $V_{FP}= 288 \pm 5.0 $~km/s and $V_{HM}= 332 \pm 8.7$~km s, so again slightly higher for HMF. The error in launch times were $\pm$~1h~35min (FPF) and $\pm$~1h27~min (HMF). Most importantly, the arrival time errors are  $\pm$~54min (FPF) and $\pm$~20~min (HMF), where we quoted the values for the location of \emph{STEREO-B}. Interestingly, the error is less for the HMF method by about a factor 2, which arises from the property that the arrival time correction partly compensates for the error in the arrival time due to the slightly different $\phi$.  It is important to note that the arrival time error of less than 1~hour is much shorter than the difference of $12$ hours in $t_a$ between the two methods (see Table~1). It is clear that these errors are rather small because we used a very long track which reaches completely up to the elongation of the spacecraft in the Jmap. We will test the accuracies and errors of shorter tracks in Section~\ref{sec:accuracy_length}, but before we check the results obtained in this section  against independent multi-point in situ data.

\subsection{STEREO-B,  \emph{VEX} and ACE in situ observations}

\begin{figure}
\epsscale{.70}
\plotone{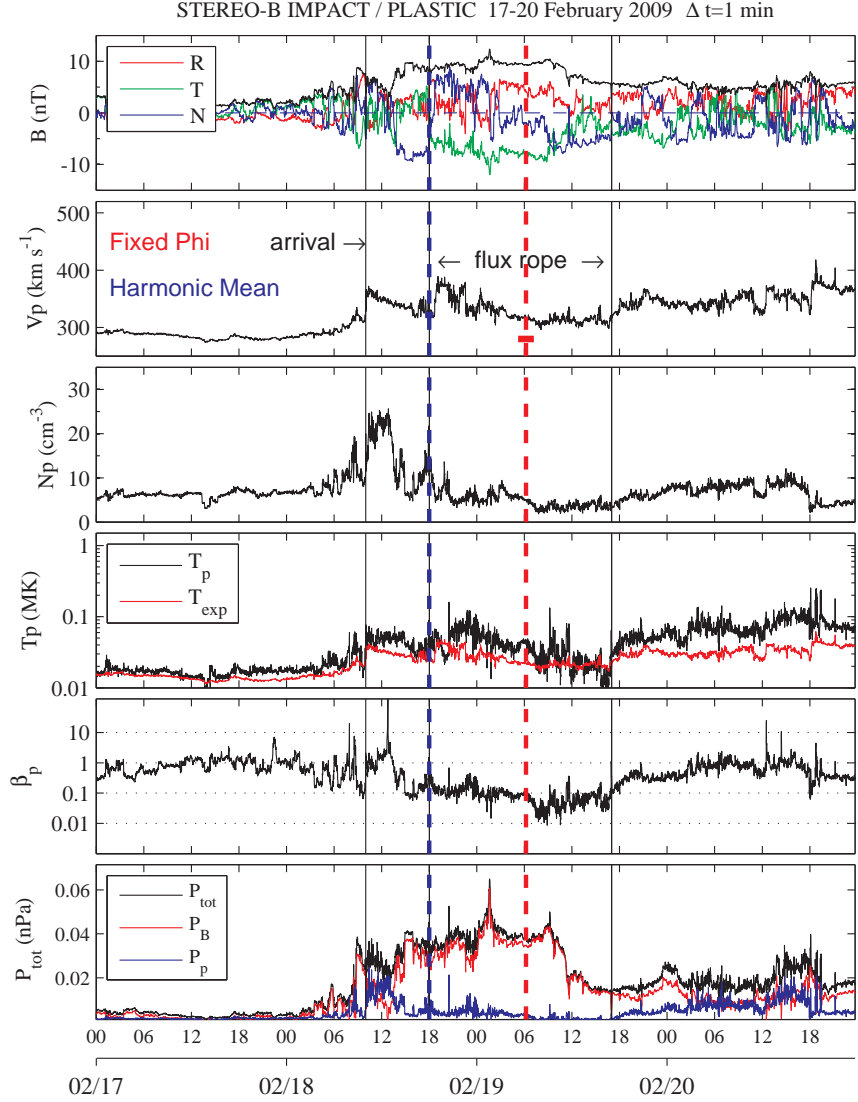}
\caption{Magnetic field and plasma data for \emph{STEREO-B}. The first solid line marks the approximate arrival time of the density front. The 2nd and 3rd solid lines indicate the flux rope interval. The arrival times from the FPF and HMF methods are given by red and blue vertical dashed lines, respectively, and the fit velocities $V_{FP}$ and $V_{HMI}$ are given by red and blue horizontal lines, which indicate the extent of the arrival time error $\pm 54$~min (FPF) and $\pm 20$~min, which are derived from the complete ICME track. From top to bottom: magnetic field magnitude and components in RTN coordinates (R pointing radially away from the Sun, T is the cross product of the solar rotation axis and R, N completes the right-handed triad),  proton bulk velocity, proton number density,  proton temperature (black) and expected temperature (red), proton $\beta$ and the total, magnetic and plasma pressure.}\label{fig:stbinsitu}
\end{figure}

The\emph{ STEREO-B} spacecraft was situated at -91 degrees (HEE longitude) to \emph{STEREO-A} during our event. This is $\Delta_B=-9^{\circ}$ and $\Delta_B=+20^{\circ}$ away from the ICME directions given by FPF and HMF, respectively. Thus, the directions from both methods indicate that the ICME will hit \emph{STEREO-B}. Figure~\ref{fig:stbinsitu} shows the proton bulk parameters \citep{gal08} and the magnetic field \citep{luh08} measured in situ. Because of its low velocity, the ICME does not drive a clear shock, but between 2009 Feb 18 08:00--10:00 UT two shock-like features mark the transition from very slow solar wind ($300$~\kms) to an enhanced density region traveling with  $350-360$~\kms. We thus define the arrival time at \emph{STEREO-B} $t_B$= Feb 18 10:00 UT, which marks the beginning of the proton density enhancement  (first vertical guideline). Because we tracked the leading edge of the ICME in the Jmap, also related to the most enhanced densities, this is the time which has to be used for assessing the arrival time predictions. After what could be a small-scale flux rope a discontinuity around Feb~18~17:00~UT marks the beginning of a large-scale flux rope (diameter 0.186~ AU), which is not a magnetic cloud \citep{bur81} because the magnetic field is not smooth and the temperature is higher compared to its environment. Nevertheless, visually it can be perceived that the field rotates as the  $B_N$ component (in this case the poloidal one) goes from north to south and $B_T$ (the axial component) is negative or eastward, throughout the flux rope interval. Assuming this rotation to be caused by the spacecraft going through a helical flux rope, it is of north--south--east (NES) type and of right-handed chirality according to the classification scheme for bipolar magnetic clouds by \cite{bot98}. We conclude that the in situ magnetic field and plasma parameters at \emph{STEREO-B} are in rough agreement with what is seen in the HI images: a slow ICME with an axis inclination $\theta$ close to the ecliptic, and hit slightly off-center because of its northward propagation direction.

In Figure~\ref{fig:stbinsitu} we also plotted the arrival times of the HMF method (vertical dashed blue line) with the velocity $V_{HMI}$ (horizontal blue line), and the same has been done for the FPF results in red. The difference between the predicted arrival time by HMF and the actual arrival time at \emph{STEREO-B} is $t_{aHM}-t_B=+8$~hours, so positive values indicate arrival times later than predicted. For, FPF, $t_{aFP}-t_B=+20$~hours, so the predicted arrival is almost a day later than the observed one. Here, the new HMF method including the arrival time correction would have given a much better prediction. The difference to the actual arrival time is also well outside of the error bars which arise from the manual selection of the ICME front. We thus attribute the residual discrepancy mainly due to the ICME leading edge geometry, which could clearly be different from the Harmonic Mean circle, and associated line-of-sight effects. Nevertheless, the point is that the new HMF method is more accurate by about 12~hours for this particular ICME than the original FPF method.

\begin{figure}
\epsscale{.95}
\plotone{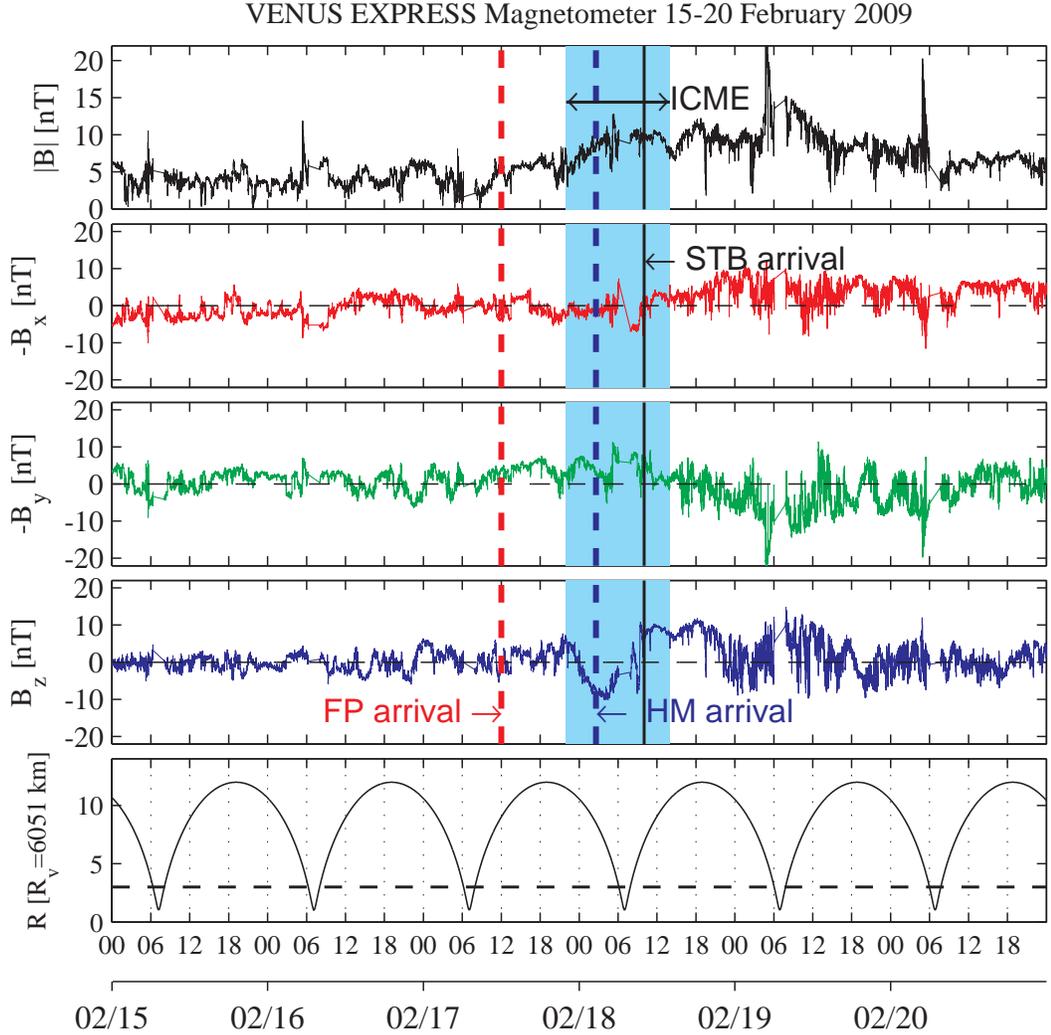}
\caption{Venus Express magnetometer observations on 2009 February 15--20. Originally, the magnetic field components are given in Venus solar orbital coordinates, where X is directed towards the Sun, Y is opposite to the planets orbital velocity ($\approx$ solar east), and Z points towards north completing the right handed triad. To make a comparison easy with Figure~\ref{fig:stbinsitu}, we plotted here $-B_x$ and $-B_y$ components so that they approximately match the orientation of the RTN coordinate system (though for the position of Venus)  used for the \emph{STEREO-B} magnetic field components. The arrival times at Venus are marked again by red (FPF) and blue (HMF) dashed lines. An ICME interval of higher than average magnetic field strengths and smoothly rotating field from Feb 17 22:00 UT -- Feb 18 14:00 UT is highlighted (shaded). For comparison, the arrival time at \emph{STEREO-B} is indicated by a solid vertical line.
 }\label{fig:VEXinsitu}
\end{figure}

Venus Express (\emph{VEX}) was positioned $-69^{\circ}$ away from \emph{STEREO-A} in longitude, so $\Delta_V=+13^{\circ}$ and $\Delta_V=+42^{\circ}$ away from the ICME directions given by FPF and HMF, respectively. While one will expect the ICME to hit  \emph{VEX} from the FPF method, with a direction between \emph{STEREO-B} and  \emph{VEX}, it is not immediately clear if the HMF direction indicates a hit, though, as already discussed before, the HMF geometry, if taken literally, will predict a hit at any spacecraft in the half space centered on $\phi_{HM}$. The arrival time correction for Venus Express is already about 30 hours at this longitudinal separation. In Figure~\ref{fig:VEXinsitu} we show \textbf{B} measured by  \emph{VEX} in Venus solar orbital coordinates \citep{zha06}, but rotated by 180 degrees ($B_x=-B_{xVSO}; B_y=-B_{yVSO}$) so it approximately matches the RTN coordinates orientation we used for \emph{STEREO-B}. We linearly interpolated over time intervals when  \emph{VEX} was below 3 Venus radii to see only the solar wind magnetic field. A magnetic flux rope interval, with a smooth rotation during an elevated field strength of about 10 nT, is found to be starting at 2009 Feb 17 22:00 UT, which we define as the Venus arrival time $t_V$.   The interval continues until about 2009 Feb 18 14:00 UT. Qualitatively, its field is oriented mostly southward and westward, but not clearly rotating and also not consistent with the field rotation at \emph{STEREO-B}. However, we think it is very unlikely that this is connected to a different ICME event, because the ICME is seen passing over the location of Venus, which is very close to the location of \emph{STEREO-B} along the line-of-sight, in the online movie around Feb~18. Also, in latitude, the ICME is expected to hit Venus centrally. This puzzling inconsistency in the field could point out that the part of the flux rope where the field is most strongly wound and which carries the major part of the magnetic flux and helicity away from the Sun does not extend over its full length back to the Sun, but  is rather confined around the apex part of the ICME \citep[recently also discussed by][]{yam10,moe10,kah11}. The error in the arrival times for FPF was $t_{aFP}-t_V \approx -10$~hours and for HMF $t_{aHM}-t_V \approx +4$~hours, which means that again the consistency with the HMF method is better. However, we note that without high time resolution data for solar wind plasma parameters it is rather difficult to set the arrival time unambigously and the arrival of the ICME is less well defined as for \emph{STEREO-B}.

\begin{figure}
\epsscale{.85}
\plotone{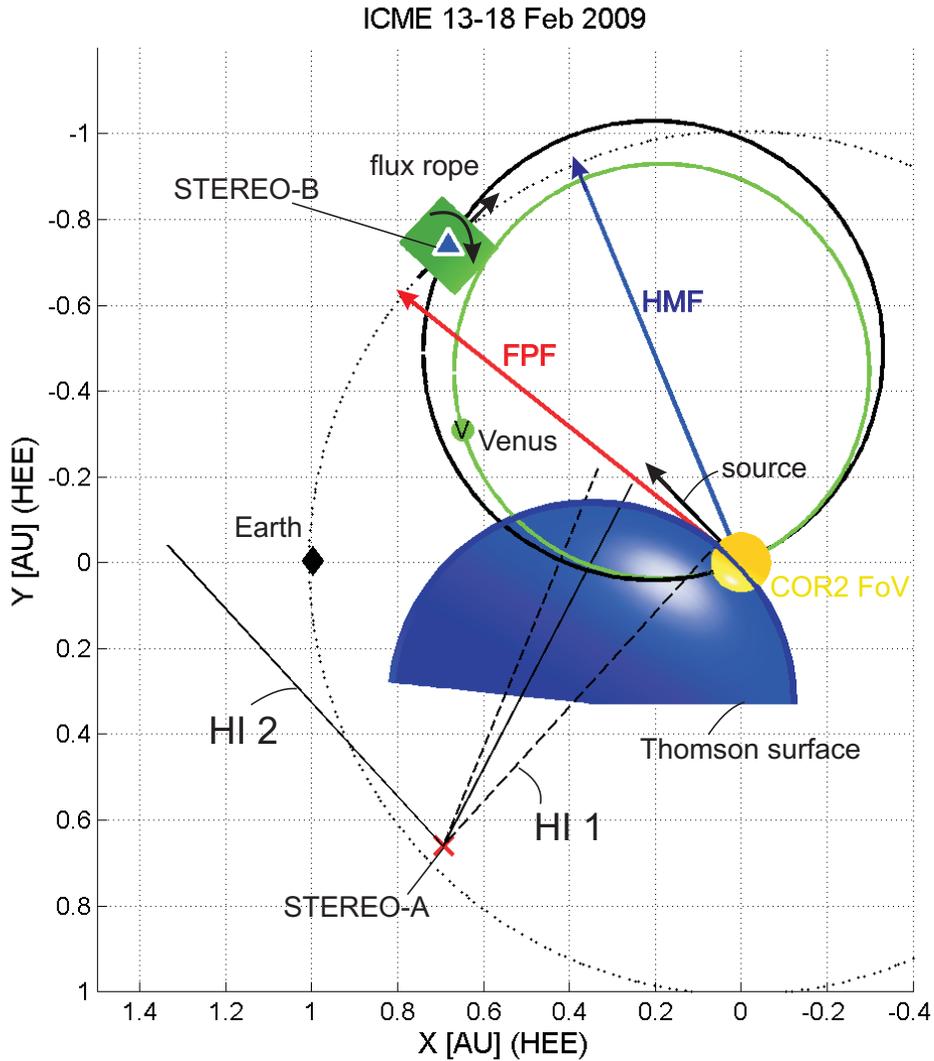}
\caption{Illustration of the ICME directions inferred for the 2009 February 13--18 ICME event using the FPF and HMF methods, looking from north down onto the ecliptic plane. The positions of the \emph{STEREO} spacecraft, Sun and Earth are indicated. The black (green) circles gives the shape of the Harmonic Mean front at the arrival time of \emph{STEREO-B} (Venus). Because the HMF method assumes a circular front, the calculated arrival times at both \emph{STEREO-B} and Venus matches better the observed arrival times. Two black arrows show the rotation of the magnetic field inside the flux rope at \emph{STEREO-B}, its radial size indicated by a green cylinder.} \label{fig:3d}
\end{figure}

In Figure~\ref{fig:3d} we plotted the event geometry and visualized the ICME directions and spacecraft positions. It can be seen that the circular front assumed by HMF leads to a \emph{negative} shift in $t_a$ compared to FPF at the position of \emph{STEREO-B} and to a \emph{positive} shift in $t_a$ at the position of  \emph{VEX}, as a direct consequence of the different geometries. Both time shifts from the FPF to the HMF method are more consistent with the in situ arrival times as discussed above. The improvement in arrival time consistency is 10 hours at the position of \emph{STEREO-B} and 4 hours at  \emph{VEX}. It is seen in Figure~\ref{fig:3d} that this arises from the circular front which touches the \emph{STEREO-B} and \emph{VEX} positions, despite the very different radial distances, at almost the same time (see also Figure~\ref{fig:VEXinsitu}). 

The Earth was $-38.5^{\circ}$ and $-67^{\circ}$ away from the ICME directions given by FPF and HMF, respectively, at a heliospheric distance of 0.9877~AU. While FPF gives an arrival time of 2009~Feb~18 22:53~UT at the \emph{ACE}/Wind satellites, a value which is quite close to the arrival time $t_B$ at \emph{STEREO-B} because of the almost similar heliospheric distance, the arrival time given by HMF is 7.5 days later on 2009~Feb~26~07:49 UT, which is due to a large arrival time correction because of the large longitudinal separation between Earth and the HM apex. We omit the plot showing the in situ data by the \emph{ACE}/Wind spacecraft near Earth because there are no clear signatures of an ICME or any large-scale magnetic flux rope between Feb 18--26.  This means that a longitudinal separation of 40--70 degrees is too far away to detect the flux rope in this case, even if it has a low axis inclination. If we assume the ICME apex at 1 AU is close to \emph{STEREO-B} (so the ICME travels radially outward), and the ICME has an axis inclination to the ecliptic $\theta\approx  0$, then the non-detection near Earth $47^{\circ}$ west of the apex  leads to the constraint that the longitudinal width of this slow solar-minimum ICME must be  $< 94^{\circ}$, which is roughly consistent with the value of $\approx 70^{\circ}$ \citep{woo10} for a different slow solar-minimum ICME event observed by STEREO/HI, and the $ < 60^{\circ}$ extent obtained from multi-spacecraft in situ observations by \cite{bot98}.


\subsection{Accuracy as a function of track length}
\label{sec:accuracy_length}
\begin{figure}
\epsscale{.75}
\plotone{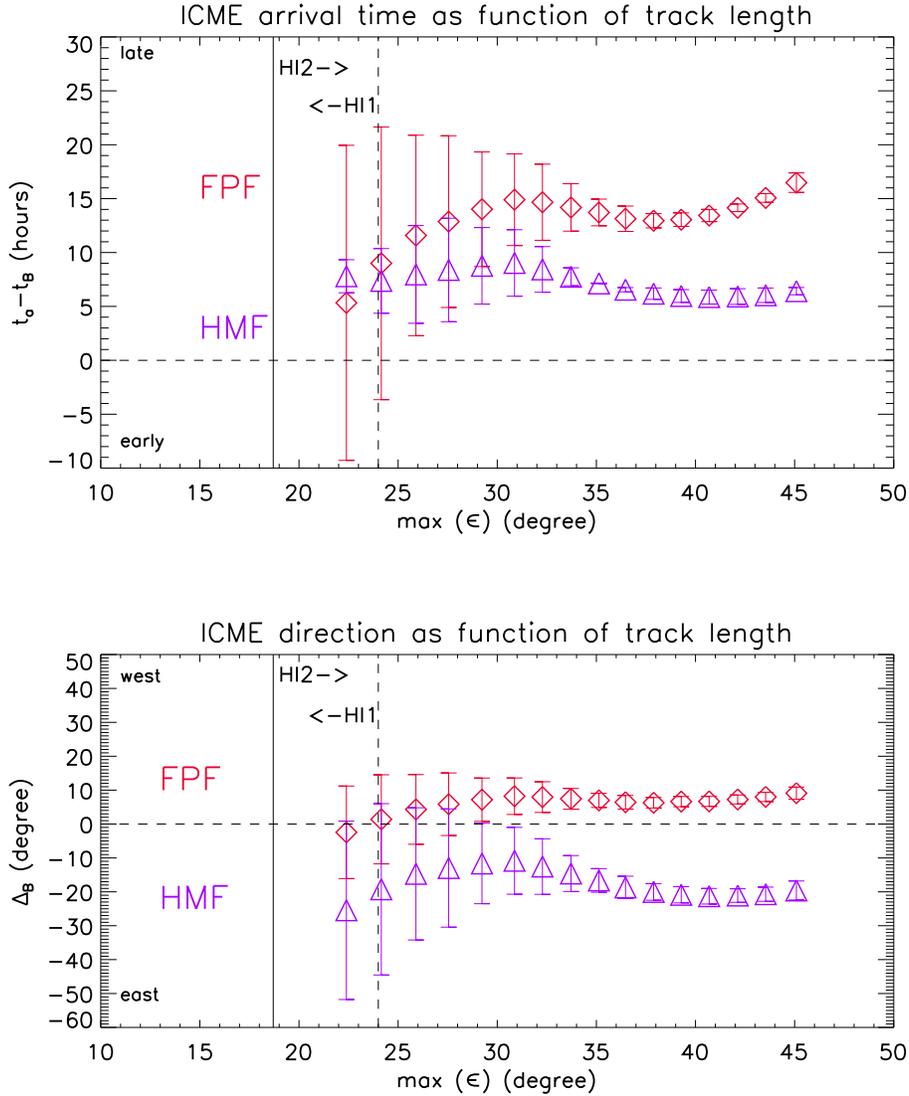}
\caption{Accuracy of possible predictions for the ICME arrival time (upper panel) and direction (lower panel) with respect to the arrival time and location of \emph{STEREO-B}, using different lengths for the tracks extracted from the Jmap, given by their maximum elongation value max$(\epsilon)$. \emph{Top:} The difference of the arrival times $t_a$ calculated for the FPF (red) and HMF (blue) methods to the observed arrival time $t_B$ at \emph{STEREO-B}. Indicated are also the outer boundary for the HI1 FoV (dashed vertical line) and the inner boundary of the HI2 FoV (solid vertical line).
\emph{Bottom:} Differences between the calculated propagation directions $\phi$ and the longitude of \emph{STEREO-B} ($\Delta_B=\phi-l_B$), as function of  max$(\epsilon)$. }
\label{fig:accuracy_length}
\end{figure}

 So far we have focused on the systematic differences which arise from the different geometric assumptions underlying the two techniques. But a major goal of future missions concerned with  real-time space weather forecasting should be to have the longest possible prediction lead times. Commonly it is quoted that the ICME must be observed to at least 30 degrees from the Sun \citep[e.g.][]{wil09,lug10} to get reliable results with HMF and FPF. To check this statement for our case study, we plotted in Figure~\ref{fig:accuracy_length}a the error in the arrival time at \emph{STEREO-B} as a function of the angle of the maximum $\epsilon$ measurement taken as an input for the fitting methods. 
We also include the error bars for $t_a$ described in Section~\ref{errorsection}. If only the measurements from the HI1 instrument are used between $\epsilon=4-24^{\circ}$, the resulting errors in the arrival time are  $\pm 12.5$~hours, so basically a day, for FPF, but much less for HMF, due to the compensation by the arrival time correction mentioned above. As expected, for higher values of max$(\epsilon)$ the errors, in particular for FPF become less and less, and are $\pm 4.8$~hours for max$(\epsilon)=30^{\circ}$, corresponding to a prediction lead time of about 48 hours until the actual arrival  for this slow ICME. Also, for max$(\epsilon)> 30^{\circ}$, the predictions by FPF are always delayed with respect to the HMF values by $\approx 10$ hours, thus, the systematic difference between the arrival times given by the two methods is not affected by the choice of max$(\epsilon)$.

In Figure~\ref{fig:accuracy_length}b, we plotted in a similar manner the difference between the heliospheric longitude of \emph{STEREO-B} and the propagation directions as results of the fitting methods for increasing values of max$(\epsilon)$. We do not know the exact propagation direction of the ICME, but looking at the error bars, their size becomes clearly more reasonable for max$(\epsilon)>30$\degree, where all directions range between
$\Delta_B=[-20,10]$\degree,  and they would all have predicted a hit at the \emph{STEREO-B} spacecraft. Also, for all times, the HMF method gives a direction further away from the observer than FPF, and the size of the error bars evolves similar in time.

\section{Discussion and conclusions}

 We discussed several new elements concerning the relationship between heliospheric images of interplanetary coronal mass ejections  obtained by \emph{STEREO/HI} and their signatures observed in situ.  We subjected two methods to forecast the ICME parameters direction, speed and arrival time to close scrutiny. These methods approximate the ICME front by a point (Fixed-$\Phi$) or a circle (Harmonic Mean), and the observed elongation of the ICME front as a function of time is fitted with inverted formulas to derive these parameters.  We introduced (1) an alternative, simpler version for the inverted formula for the Harmonic Mean fitting method, and  (2) an analytic arrival time correction formula which is geometrically consistent with an assumed circular front shape of the ICME. It should be applied to any real time predictions or case and statistical studies which discuss ICME arrival time calculation using methods which assume a circular ICME front, i.e.\ single-spacecraft Harmonic Mean Fitting \citep{lug10b} or two-spacecraft tangent-to-a-sphere \citep{lug10}. 

We then checked the validity of the derived formulas by applying them in a case study to the 2009 February 13--18 ICME event, a perfect quadrature event, seen at the limb by \emph{STEREO-A} and in situ by \emph{STEREO-B} and Venus Express. The event was a slow event, with approximately constant speed throughout the HI field of views. We found that the arrival time given by the Harmonic Mean Fitting (HMF) method was more accurate than the  Fixed-$\Phi$ fitting (FPF) method by 12 hours, in direct comparison to the observed in situ  arrival time at \emph{STEREO-B}. We attribute this to the fact that for an ICME with low axis inclination $\theta$ to the ecliptic, the ICME front is better described by the circular geometry assumed by Harmonic Mean. The low $\theta$ value was inferred independently by the morphology and small de-projected angular width of the ICME in \emph{STEREO-A/HI} and the magnetic field rotation observed by \emph{STEREO-B}. At the latter spacecraft, the rough behavior of the magnetic field components pointed to a slight off-center hit of the ICME, consistent with a northward propagation by a few degrees of the ICME seen in HI. Thus, \emph{STEREO-B} crossed the outer part of the cross-section of the flux rope. 
These observations are also consistent with a view that the outer part of the flux rope is classified as an ejecta and not a magnetic cloud, with a rather flat pressure profile \citep{jia06}. Additionally, Venus Express observed a magnetic flux rope too, about 23 degrees west of STEREO-B,  and again the in situ arrival time was more consistent with the one predicted by HMF. We also assessed the errors in the arrival time predictions as a function of how long the ICME is visible in the heliospheric images, and found that one needs to track the ICME up to 30 degrees in elongation to obtain systematic errors (arising from the manual selection of the track only) which are less than $\pm 5$ hours when using scientific (level 2) HI data.

It is our new hypothesis in this paper that the two methods for fitting time-elongation data  (HMF and FPF) are just limiting cases of ICMEs with low or high axis inclination to the ecliptic, using two ``extreme'' assumptions on the ICME front shape (circular vs. point-like). The same can also be said for the triangulation \citep[point-like,][]{liu10} and tangent-to-a-sphere \citep[circular,][]{lug10} methods, which use two-spacecraft HI observations. In our hypothesis, the HMF method corresponds to the case of low axis inclination and thus a large longitudinal extent in the ecliptic \citep[e.g.\ 70 degrees for a solar minimum low-inclination ICME,][]{woo10} along the line-of-sight. The FPF method should give better results for an ICME with a high axis inclination because the extent of the cross-section is generally thought to be much smaller along the line of sight than its longitudinal extension \citep[e.g.\ about 20 degrees for a solar minimum high-inclination ICME, see][]{kil09a,moe09b,moe09}.  

This event was especially well suited for testing this hypothesis because the constant speed and constant direction assumptions of the HMF and FPF methods were approximately fulfilled,  thus the systematic differences we found should arise only from the truly different shape of the ICME front. In reality, these simplified geometric assumptions will only be rough approximations the ICME front shape, and for slow events, which tend to get swept up in corotating interaction regions and become more easily distorted, any simple geometric model might be too restrictive in general.  But for faster and more geo-effective events, where the front shape might be expected to be less easily affected by the background solar wind structure, the fitting methods could provide increasingly accurate results.

We plan to extend this type of study for a larger set of events to find out how feasible these techniques are, if they have to be extended to other geometries of the ICME front and to further test our hypothesis concerning the differences in axis inclination. Also, one of the major questions will be if one or rather two spacecraft are necessary for a L5 or a combined L4/L5 mission \citep[e.g.][]{liu10b, gop11} to provide us with sufficient accuracy for advance warnings of solar eruptions using heliospheric imagers. The methods described in this paper can be used for such a mission, and also for any upcoming space weather forecasts at Earth and other planets with \emph{STEREO}. \cite{dav11} showed the efficacy of using the FPF method for real-time prediction of an ICME event in April 2010, when the \emph{STEREO} spacecraft were separated by about 70 degrees to Earth, which is comparable to the L5 point at 60 degrees. There is still a long way to go, but with a solar maximum with many ICMEs in front of us we will be able to use \emph{STEREO} extensively to continue honing our skills.

\begin{table}
\begin{center}
\caption{A compilation of observed parameters and results of fitting techniques for the 2009 February 13--18 ICME. The results for COR1A are taken from Miklenic et al. (2011, submitted to Solar Physics). FPF: Fixed-$\phi$ fitting; HMF: harmonic mean fitting; The suffix ``/B'' indicates results for \emph{STEREO-B}, ``/V'' for \emph{VEX}. Column R$_{\odot}$ indicates the range of solar radii where the measurements were obtained or techniques were applied. For the fitting methods, this is the range of the distances $R_{FP}$ and $R_{HM}$ corresponding to the first and last $\epsilon(t)$ measurement of the ICME track in the Jmap. Angle $l$ is the longitude of the source region (flare peak), the ICME direction (FPF/HMF) and the spacecraft positions (\emph{STEREO-B} / \emph{VEX}) with respect to the \emph{STEREO-A} spacecraft ($l> 0$ for solar west). $V$ is the CME/ICME speed, and $t_0$ /  $t_a$ are the launch and arrival times, respectively.
\label{tab:cmedir} }
\begin{tabular}{ccccccc}
\tableline\tableline
technique      & Instruments & $R_{\odot}$& $l$, deg &  V,  km s$^{-1}$ & $t_0$, UT &  $t_a$, UT \\
\tableline
 flare peak  & GOES & 1      & $-89.5$       & -          & Feb 13 05:46 & -                     \\
tracking & COR1A     & 2.5  & $-91$\tablenotemark{a}   & 350\tablenotemark{a}        & Feb 13 06:20     & Feb 18 03:57\tablenotemark{b}    \\
\tableline
FPF/B & HI1/2A    & 14.1--203.7 & $-79$  & 280  & Feb 13 01:30 & 19 Feb 06:12   \\
HMF/B & HI1/2A    & 14.7--225.7 & $-107$ & 322 / 310\tablenotemark{c}& Feb 13 03:20 & 18 Feb 18:00  \\
STB & IMPACT/PLASTIC & 215.6 & $-91$          & 362 / 348\tablenotemark{d}& -  & 18 Feb 10:00\\
\tableline
FPF/V & HI1/2A    & 14.1--203.7 & $-79$ & 280  & Feb 13 01:30 & 17 Feb 12:01   \\
HMF/V & HI1/2A    & 14.7--225.7 & $-107$ & 322  & Feb 13 03:20 & 18 Feb 02:36  \\
 VEX & MAG &   154.5 &  $-68.9$ & - & -  & 17 Feb 22:00 \\
\tableline
\end{tabular}
\end{center}
 \tablenotetext{a}{The COR1A speed was measured in the plane-of-sky by Miklenic et al.\ (2011, submitted to Solar Physics).}
 \tablenotetext{b}{Ballistic projection of the CMEs arrival at \emph{STEREO-B} with the COR1 measurements. }
\tablenotetext{c}{The first speed is for the apex of the HM circle ($V_{HM}$) and the second for the flank part hitting \emph{STEREO-B}  ($V_{HMI}$).  } 
 \tablenotetext{d}{The in situ $V$ are means over the ICME sheath / flux rope regions.  }
%
\end{table}



%


\acknowledgments

C.M., M.T. and T.R. were supported by the Austrian Science Fund (FWF): P20145-N16, V195-N16. The presented work has received funding from the European Union Seventh Framework Programme (FP7/2007-2013) under grant agreement n$^{\circ}$263252 [COMESEP].  It is also supported by NASA grants  NAS5-00132, NNG06GD41G, NNX08AD11G, NNX10AQ29G and  NNX08AQ16G. Work at the University of California, Berkeley, was supported from STEREO grant NAS5-03131. The SECCHI data are produced by an international consortium of Naval Research Laboratory, Lockheed Martin Solar and Astrophysics Lab, and NASA Goddard Space Flight Center (USA), Rutherford Appleton Laboratory, and University of Birmingham (UK), Max-Planck-Institut f\"ur Sonnensystemforschung (Germany), Centre Spatiale de Liege (Belgium), Institut d'Optique Theorique et Appliquee, and Institut d'Astrophysique Spatiale (France).


\newpage

\appendix

\section{Alternative version of the inverted Harmonic Mean formula}

\begin{figure}
\epsscale{.70}
\plotone{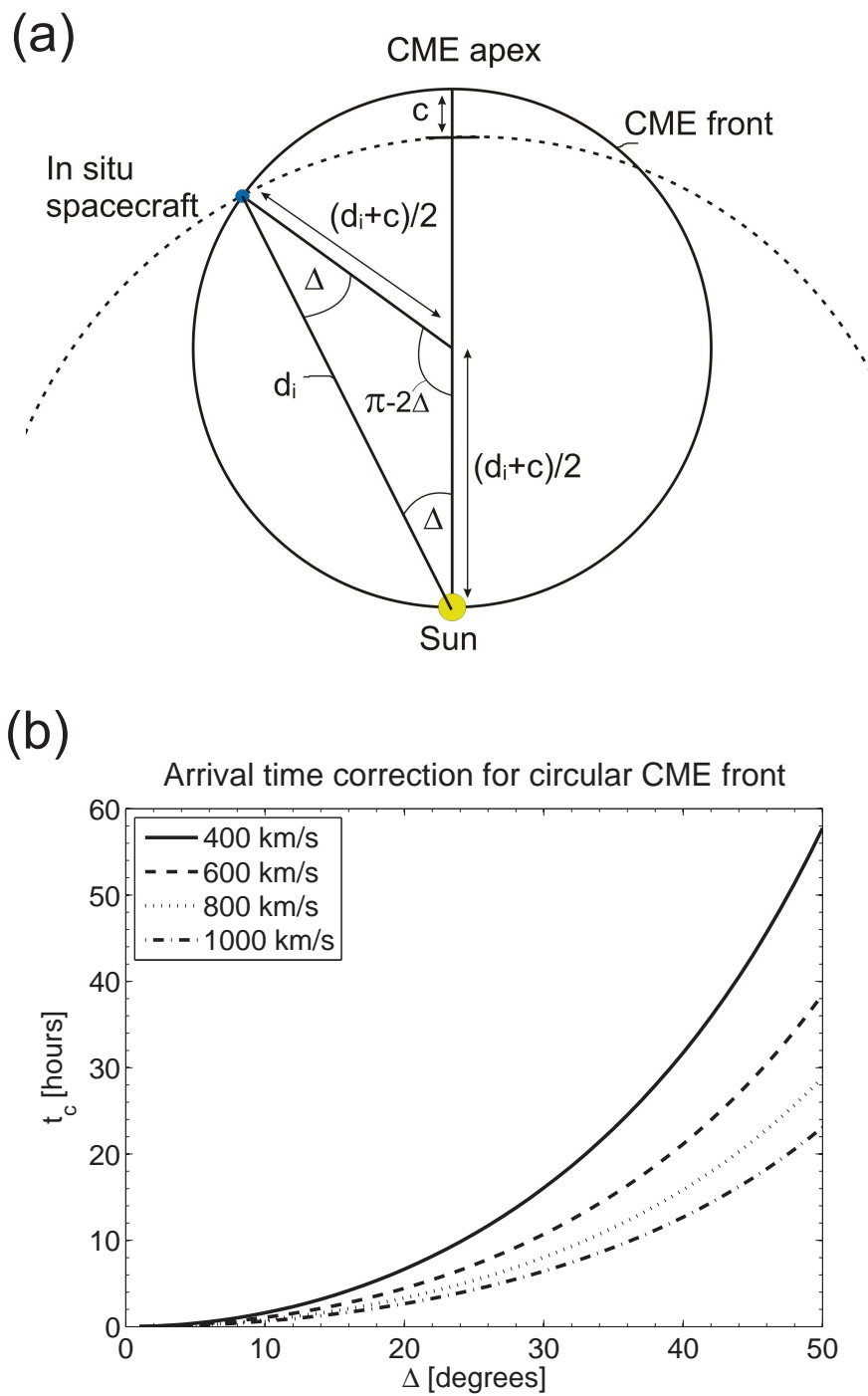}
\caption{(a) A sketch showing the definitions and geometry for deriving the arrival time correction for the HMF and TAS methods (see text). The solid circle is the circular front assumed for the ICME leading edge and the dashed line depicts 1~AU. (b) The arrival time correction -- Equation (\ref{equ:arrtimehm_niceunits}) -- is plotted as a function of the angular separation between the ICME apex and the in situ observing  spacecraft $I$, which is situated at a radial distance from the Sun of $d_i=1$~AU.}\label{fig:correction}
\end{figure}

The aim of this section is to explicitly show how to solve Equation~(\ref{equ:hmconversion}) for $\epsilon(t)$. This is the conversion formula from elongation to distance if one assumes that the observer looks along a tangent to a circle which is always attached to the Sun at one end (``Harmonic Mean" assumption, see Figures~1 and \ref{fig:correction}a). The radial distance of the circle's apex $R_{HM}$, the point with largest distance from the Sun, is given by \cite{lug09a}:

\begin{equation}\label{equ:hmconversion_back}
    R_{HM}(t)=  \frac{2 d_o \sin \epsilon(t)}{1+\sin(\epsilon(t)+\phi)},   \\
\end{equation}
with $d_o$ the radial distance of the observer from the Sun, $\epsilon(t)$ the observed elongation as a function of time and $\phi$ the constant ICME propagation direction with respect to the observer ($\phi > 0$ means solar west).
After some basic algebra, Equation~(\ref{equ:hmconversion_back}) can be written as 
 \begin{equation}
    \left(\frac{2d_o}{V_{HM} t}-\cos{\phi_{HM}} \right ) \sin (\epsilon (t)) - \sin(\phi_{HM}) \cos(\epsilon(t))=1, \\
\end{equation}
where we have already replaced distance with speed: $R_{HM}=V_{HM}t$. For better readability, we define
\begin{equation}
   a=\frac{2d_o}{V_{HM} t}-\cos(\phi_{HM}); \quad   b=\sin(\phi_{HM}), \\
\end{equation}
which results in 
 \begin{equation}
    a \sin (\epsilon (t)) - b \cos(\epsilon (t))=1. \\
\end{equation}
This equation can be brought into quadratic form by re-arranging it to 
 \begin{equation}
    a \sqrt{1- \cos^2 (\epsilon (t))}=1+ b \cos(\epsilon (t)) \\
\end{equation}
and squaring both sides. Solving the resulting quadratic equation in  $\cos{(\epsilon(t))}$ leads to our form of the Harmonic Mean fitting equation as
\begin{equation}\label{equ:hmfit}
   \epsilon (t) = \arccos ( \frac{-b + a \sqrt{a^2+b^2-1}}{a^2+b^2}).  \\
\end{equation}

 We conclude this section with some  notes useful when developing a code that fits the inverse equations. While the $\arctan$ function in the FPF Equation (\ref{equ:fpfit_vorn}) is symmetric with respect to positive and negative values of the direction $\phi$,  the HMF equation (\ref{equ:hmfit}) will always produce positive values for $\epsilon(t)$ because of the $\arccos$ function, regardless of the sign of the direction $\phi_{FP}$. Elongations provided by \emph{STEREO-A}/HI, which looks to solar east, are negative by definition, so these $\epsilon < 0$  directly translate to $\phi_{FP}< 0$ (or solar east) when applying the FPF method. For \emph{STEREO-B} , automatically $\epsilon > 0$ and thus  $\phi_{FP}> 0$, consistent with our definition that   $\phi_{FP}> 0$ corresponds to solar west. In contrast, when applying the HMF method to \emph{STEREO-A}, always positive elongations have to be fitted and the sign of the direction $\phi_{HM}$ has to be additionally corrected. The reason for this asymmetry is that the observer is also able to measure an ICME front with HMF even if the apex of the HM circle is on the other side of where the Heliospheric imager is looking, i.e.\ when the apex is west of \emph{STEREO-A} or east of \emph{STEREO-B} \citep{lug10b}.

\section{Arrival time calculation for circular ICME fronts}

For calculating the arrival time of a circular ICME front at a given position in the heliosphere, as assumed by the Harmonic Mean (HM) fitting \citep{lug10b} or the tangent-to-a-sphere \citep{lug10} methods, it is necessary to derive a formula which takes into account that different parts of the front will move with different speeds. As a consequence, the ICME arrival time will be a function not only of the radial distance of the in situ spacecraft $I$ (or planet) from the Sun but also of its separation angle $\Delta$ to the HM apex, the point with the largest distance from the Sun.

 If the ICME in question is not heading exactly towards the in situ observing spacecraft $I$, i.e.\ the ICME direction $\phi_{HM}$ as the result of the Harmonic Mean fitting method is not the same as the longitudinal separation between the observer $O$ and $I$, there will be a time delay $t_c \geq 0$ when the ICME hits $I$ compared to when the ICME apex is at  the same radial distance of $I$ from the Sun. Figure \ref{fig:correction}a visualizes this geometry. We will first derive this formula as a ``correction'' $t_c$ compared to the ICME apex to assess its significance, and will then derive the formula useful for direct calculation of the arrival time $t_{aHM}$. The arrival time correction $t_c$ is a function of the angular separation $\Delta=\phi_{HM}-l_i$ between $\phi_{HM}$ and the heliospheric longitude of $I$ (called $l_i$), the velocity $V_{HM}$ obtained by the fitting procedure and the radial distance $d_i$ of the in situ spacecraft from the Sun: $t_c(\Delta, V_{HM},d_i)$. We use all longitudinal angles as such that westward is defined as positive, and the condition $\Delta \in [-\pi/2 ... +\pi/2]$ is necessary for the spacecraft to be hit by the expanding circular front. Additionally,  $c$ is defined as the distance the HM apex travels before the HM circle hits the spacecraft $I$. Applying the law of sines to the triangle indicated in Figure \ref{fig:correction}a, we find
\begin{gather}
    \frac{(d_i+c)/2}{\sin(\Delta)}=\frac{d_i}{\sin(\pi-2\Delta)}.
\end{gather}
Simple regrouping leads to
\begin{equation}
    c=d_i\left (\frac{1}{\cos(\Delta)}-1 \right).
\end{equation}
To convert this distance to a time, we assume that the ICME continues to travel with the constant velocity
obtained from the fitting procedure ($V_{HM}$) and define $t_c \geq 0$ as the arrival time correction,
\begin{gather}
    t_{c}=c/V_{HM}\\
    t_{c}=\frac{d_i(\cos^{-1}(\Delta)-1)}{V_{HM}}\label{equ:arrtimehm_normalunits}
  \end{gather}
This can be rewritten as the simple formula
\begin{equation}\label{equ:arrtimehm_niceunits}
    t_{c}[\rm{h}]=41555 \times \frac{\it{d_i}[\rm{AU}](\cos^{-1}(\Delta)-1)}{ \it{V_{HM}} [\rm{km~s^{-1}}]}
\end{equation}
with the result $t_c$ in hours and $V_{HM}$ and $d_i$ given in convenient units. Figure \ref{fig:correction}b shows the behavior of $t_c(\Delta)$ for $d_i=1$~AU and various $V_{HM}$. Obviously, for faster CMEs the arrival time correction becomes less, and for a $V_{HM}=800$~km~s$^{-1}$ CME, the arrival time correction is less than 10 hours for $\Delta < 35$~degrees. It might be surprising that Equation~(\ref{equ:arrtimehm_niceunits}) depends on $d_i$ -- this is simply because the circle assumed by HM is always attached to the Sun at one end and thus expands while the apex propagates outward.

In summary, to calculate the arrival time of an ICME at a given heliospheric distance $d_i$ and  at a longitudinal separation $\Delta$ with the HMF method, a correction $t_c$ must be added to the arrival time of the ICME apex:

\begin{equation}\label{equ:arrtimehm1}
 t_{aHM}=t_{0HM} + \frac{d_i}{V_{HM}}+t_c,
\end{equation}
with $t_{0HM}$ being the launch time (defined here by $\epsilon(t_{0HM})=0$).
This reduces with Equation~(\ref{equ:arrtimehm_normalunits}) to simply

\begin{equation}\label{equ:arrtimehm_simple}
 t_{aHM}=t_{0HM} + \frac{d_i}{ V_{HM}\cos(\Delta)}.
\end{equation}
This is the final equation for calculating the  arrival time of an ICME with the HMF method. From the last equation, the elementary result follows that the speed of the ICME flank at a given angle $\Delta$ to the ICME apex is just reduced by a factor $\cos(\Delta)$ for a circular geometry. When comparing the speed $V_{HM}$ as an outcome of the HMF method to in situ measurements of the solar wind bulk speed, the speed $V_{HMI}=V_{HM} \cos(\Delta)$ has to be used.





\begin{thebibliography}{54}
\expandafter\ifx\csname natexlab\endcsname\relax\def\natexlab#1{#1}\fi

\bibitem[{{Bothmer} \& {Schwenn}(1998)}]{bot98}
{Bothmer}, V., \& {Schwenn}, R. 1998, Annales Geophysicae, 16, 1

\bibitem[{{Bothmer} \& {Zhukov}(2006)}]{bot06}
{Bothmer}, V., \& {Zhukov}, A. 2006, in Space Weather - Physics and Effects,
  ed. V. Bothmer and I.~A. Daglis, Springer-Verlag, 31

\bibitem[{{Burlaga} {et~al.}(1981){Burlaga}, {Sittler}, {Mariani}, \&
  {Schwenn}}]{bur81}
{Burlaga}, L., {Sittler}, E., {Mariani}, F., \& {Schwenn}, R. 1981, \jgr, 86,
  6673

\bibitem[{{Cohen} {et~al.}(2009){Cohen}, {Attrill}, {Manchester}, \&
  {Wills-Davey}}]{coh09}
{Cohen}, O., {Attrill}, G.~D.~R., {Manchester}, W.~B., \& {Wills-Davey}, M.~J.
  2009, \apj, 705, 587

\bibitem[{{Committee On The Societal} \& {Economic Impacts Of Severe Space
  Weather Events}(2008)}]{nat08}
{Committee On The Societal}, \& {Economic Impacts Of Severe Space Weather
  Events}. 2008, {Severe Space Weather Events--Understanding Societal and
  Economic Impacts: A Workshop Report, National Academies Press. ISBN:
  978-0-309-12769-1}, Tech. rep.

\bibitem[{{Davies} {et~al.}(2009){Davies}, {Harrison}, {Rouillard}, {Sheeley},
  {Perry}, {Bewsher}, {Davis}, {Eyles}, {Crothers}, \& {Brown}}]{dav09b}
{Davies}, J.~A., {et~al.} 2009, \grl, 36, 2102

\bibitem[{{Davis} {et~al.}(2009){Davis}, {Davies}, {Lockwood}, {Rouillard},
  {Eyles}, \& {Harrison}}]{dav09}
{Davis}, C.~J., {Davies}, J.~A., {Lockwood}, M., {Rouillard}, A.~P., {Eyles},
  C.~J., \& {Harrison}, R.~A. 2009, \grl, 36, 8102

\bibitem[{{Davis} {et~al.}(2010){Davis}, {Kennedy}, \& {Davies}}]{dav10}
{Davis}, C.~J., {Kennedy}, J., \& {Davies}, J.~A. 2010, \solphys, 263, 209

\bibitem[{{Davis} {et~al.}(2011){Davis}, {de Koning}, {Davies}, {Biesecker},
  {Millward}, {Dryer}, {Deehr}, {Webb}, {Schenk}, {Freeland}, {M{\"o}stl},
  {Farrugia}, \& {Odstrcil}}]{dav11}
{Davis}, C.~J., {et~al.} 2011, Space Weather, 9, 1005

\bibitem[{{Eyles} {et~al.}(2009){Eyles}, {Harrison}, {Davis}, {Waltham},
  {Shaughnessy}, {Mapson-Menard}, {Bewsher}, {Crothers}, {Davies}, {Simnett},
  {Howard}, {Moses}, {Newmark}, {Socker}, {Halain}, {Defise}, {Mazy}, \&
  {Rochus}}]{eyl09}
{Eyles}, C.~J., {et~al.} 2009, \solphys, 254, 387

\bibitem[{{Farrugia} {et~al.}(1993){Farrugia}, {Burlaga}, {Osherovich},
  {Richardson}, {Freeman}, {Lepping}, \& {Lazarus}}]{far93}
{Farrugia}, C.~J., {Burlaga}, L.~F., {Osherovich}, V.~A., {Richardson}, I.~G.,
  {Freeman}, M.~P., {Lepping}, R.~P., \& {Lazarus}, A.~J. 1993, \jgr, 98, 7621

\bibitem[{{Galvin} {et~al.}(2008){Galvin}, {Kistler}, {Popecki}, {Farrugia},
  {Simunac}, {Ellis}, {M{\"o}bius}, {Lee}, {Boehm}, {Carroll}, {Crawshaw},
  {Conti}, {Demaine}, {Ellis}, {Gaidos}, {Googins}, {Granoff}, {Gustafson},
  {Heirtzler}, {King}, {Knauss}, {Levasseur}, {Longworth}, {Singer}, {Turco},
  {Vachon}, {Vosbury}, {Widholm}, {Blush}, {Karrer}, {Bochsler}, {Daoudi},
  {Etter}, {Fischer}, {Jost}, {Opitz}, {Sigrist}, {Wurz}, {Klecker}, {Ertl},
  {Seidenschwang}, {Wimmer-Schweingruber}, {Koeten}, {Thompson}, \&
  {Steinfeld}}]{gal08}
{Galvin}, A.~B., {et~al.} 2008, Space Science Reviews, 5

\bibitem[{{Gopalswamy} {et~al.}(2009){Gopalswamy}, {M{\"a}kel{\"a}}, {Xie},
  {Akiyama}, \& {Yashiro}}]{gop09}
{Gopalswamy}, N., {M{\"a}kel{\"a}}, P., {Xie}, H., {Akiyama}, S., \& {Yashiro},
  S. 2009, Journal of Geophysical Research (Space Physics), 114, 0

\bibitem[{{Gopalswamy} {et~al.}(2011){Gopalswamy}, {Davila}, {St.~Cyr},
  {Sittler}, {Auch{\`e}re}, {Duvall}, {Hoeksema}, {Maksimovic}, {MacDowall},
  {Szabo}, \& {Collier}}]{gop11}
{Gopalswamy}, N., {et~al.} 2011, Journal of Atmospheric and Solar-Terrestrial
  Physics, 73, 658

\bibitem[{{Gosling}(1993)}]{gos93}
{Gosling}, J.~T. 1993, \jgr, 98, 18937

\bibitem[{{Howard} {et~al.}(1985){Howard}, {Sheeley}, {Michels}, \&
  {Koomen}}]{how85}
{Howard}, R.~A., {Sheeley}, Jr., N.~R., {Michels}, D.~J., \& {Koomen}, M.~J.
  1985, \jgr, 90, 8173

\bibitem[{{Howard} {et~al.}(2008){Howard}, {Moses}, {Vourlidas}, {Newmark},
  {Socker}, {Plunkett}, {Korendyke}, {Cook}, {Hurley}, {Davila}, {Thompson},
  {St Cyr}, {Mentzell}, {Mehalick}, {Lemen}, {Wuelser}, {Duncan}, {Tarbell},
  {Wolfson}, {Moore}, {Harrison}, {Waltham}, {Lang}, {Davis}, {Eyles},
  {Mapson-Menard}, {Simnett}, {Halain}, {Defise}, {Mazy}, {Rochus}, {Mercier},
  {Ravet}, {Delmotte}, {Auchere}, {Delaboudiniere}, {Bothmer}, {Deutsch},
  {Wang}, {Rich}, {Cooper}, {Stephens}, {Maahs}, {Baugh}, {McMullin}, \&
  {Carter}}]{how08}
{Howard}, R.~A., {et~al.} 2008, Space Science Reviews, 136, 67

\bibitem[{{Howard} \& {Tappin}(2009{\natexlab{a}})}]{how09a}
{Howard}, T.~A., \& {Tappin}, S.~J. 2009{\natexlab{a}}, \ssr, 147, 31

\bibitem[{{Howard} \& {Tappin}(2009{\natexlab{b}})}]{how09b}
{Howard}, T.~A., \& {Tappin}, S.~J.  2009{\natexlab{b}}, \ssr, 147, 89

\bibitem[{{Jian} {et~al.}(2006){Jian}, {Russell}, {Luhmann}, \&
  {Skoug}}]{jia06}
{Jian}, L., {Russell}, C.~T., {Luhmann}, J.~G., \& {Skoug}, R.~M. 2006,
  \solphys, 239, 393

\bibitem[{{Kahler} {et~al.}(2011){Kahler}, {Krucker}, \& {Szabo}}]{kah11}
{Kahler}, S.~W., {Krucker}, S., \& {Szabo}, A. 2011, Journal of Geophysical
  Research (Space Physics), 116, A01104

\bibitem[{{Kahler} \& {Webb}(2007)}]{kah07}
{Kahler}, S.~W., \& {Webb}, D.~F. 2007, Journal of Geophysical Research (Space
  Physics), 112, 9103

\bibitem[{{Kaiser} {et~al.}(2008){Kaiser}, {Kucera}, {Davila}, {St.~Cyr},
  {Guhathakurta}, \& {Christian}}]{kai08}
{Kaiser}, M.~L., {Kucera}, T.~A., {Davila}, J.~M., {St.~Cyr}, O.~C.,
  {Guhathakurta}, M., \& {Christian}, E. 2008, Space Science Reviews, 136, 5

\bibitem[{{Kienreich} {et~al.}(2009){Kienreich}, {Temmer}, \&
  {Veronig}}]{kie09}
{Kienreich}, I.~W., {Temmer}, M., \& {Veronig}, A.~M. 2009, \apjl, 703, L118

\bibitem[{{Kilpua} {et~al.}(2009){Kilpua}, {Liewer}, {Farrugia}, {Luhmann},
  {M{\"o}stl}, {Li}, {Liu}, {Lynch}, {Russell}, {Vourlidas}, {Acuna}, {Galvin},
  {Larson}, \& {Sauvaud}}]{kil09a}
{Kilpua}, E.~K.~J., {et~al.} 2009, \solphys, 254, 325

\bibitem[{{Liu} {et~al.}(2010{\natexlab{a}}){Liu}, {Davies}, {Luhmann},
  {Vourlidas}, {Bale}, \& {Lin}}]{liu10}
{Liu}, Y., {Davies}, J.~A., {Luhmann}, J.~G., {Vourlidas}, A., {Bale}, S.~D.,
  \& {Lin}, R.~P. 2010{\natexlab{a}}, \apjl, 710, L82

\bibitem[{{Liu} {et~al.}(2010{\natexlab{b}}){Liu}, {Thernisien}, {Luhmann},
  {Vourlidas}, {Davies}, {Lin}, \& {Bale}}]{liu10b}
{Liu}, Y., {Thernisien}, A., {Luhmann}, J.~G., {Vourlidas}, A., {Davies},
  J.~A., {Lin}, R.~P., \& {Bale}, S.~D. 2010{\natexlab{b}}, \apj, 722, 1762

\bibitem[{{Lugaz}(2010)}]{lug10b}
{Lugaz}, N. 2010, \solphys, 267, 411

\bibitem[{{Lugaz} {et~al.}(2010){Lugaz}, {Hernandez-Charpak}, {Roussev},
  {Davis}, {Vourlidas}, \& {Davies}}]{lug10}
{Lugaz}, N., {Hernandez-Charpak}, J.~N., {Roussev}, I.~I., {Davis}, C.~J.,
  {Vourlidas}, A., \& {Davies}, J.~A. 2010, \apj, 715, 493

\bibitem[{{Lugaz} {et~al.}(2011){Lugaz}, {Roussev}, \& {Gombosi}}]{lug11}
{Lugaz}, N., {Roussev}, I.~I., \& {Gombosi}, T.~I. 2011, Advances in Space
  Research, 48, 292

\bibitem[{{Lugaz} {et~al.}(2009){Lugaz}, {Vourlidas}, \& {Roussev}}]{lug09a}
{Lugaz}, N., {Vourlidas}, A., \& {Roussev}, I.~I. 2009, Annales Geophysicae,
  27, 3479

\bibitem[{{Luhmann} {et~al.}(2008){Luhmann}, {Curtis}, {Schroeder}, {McCauley},
  {Lin}, {Larson}, {Bale}, {Sauvaud}, {Aoustin}, {Mewaldt}, {Cummings},
  {Stone}, {Davis}, {Cook}, {Kecman}, {Wiedenbeck}, {von Rosenvinge}, {Acuna},
  {Reichenthal}, {Shuman}, {Wortman}, {Reames}, {Mueller-Mellin}, {Kunow},
  {Mason}, {Walpole}, {Korth}, {Sanderson}, {Russell}, \& {Gosling}}]{luh08}
{Luhmann}, J.~G., {et~al.} 2008, Space Science Reviews, 136, 117

\bibitem[{{Lynch} {et~al.}(2010){Lynch}, {Li}, {Thernisien}, {Robbrecht},
  {Fisher}, {Luhmann}, \& {Vourlidas}}]{lyn10}
{Lynch}, B.~J., {Li}, Y., {Thernisien}, A.~F.~R., {Robbrecht}, E., {Fisher},
  G.~H., {Luhmann}, J.~G., \& {Vourlidas}, A. 2010, Journal of Geophysical
  Research (Space Physics), 115, 7106

\bibitem[{{M{\"o}stl} {et~al.}(2009{\natexlab{a}}){M{\"o}stl}, {Farrugia},
  {Biernat}, {Leitner}, {Kilpua}, {Galvin}, \& {Luhmann}}]{moe09b}
{M{\"o}stl}, C., {Farrugia}, C.~J., {Biernat}, H.~K., {Leitner}, M., {Kilpua},
  E.~K.~J., {Galvin}, A.~B., \& {Luhmann}, J.~G. 2009{\natexlab{a}}, \solphys,
  256, 427

\bibitem[{{M{\"o}stl} {et~al.}(2009{\natexlab{b}}){M{\"o}stl}, {Farrugia},
  {Temmer}, {Miklenic}, {Veronig}, {Galvin}, {Leitner}, \& {Biernat}}]{moe09c}
{M{\"o}stl}, C., {Farrugia}, C.~J., {Temmer}, M., {Miklenic}, C., {Veronig},
  A.~M., {Galvin}, A.~B., {Leitner}, M., \& {Biernat}, H.~K.
  2009{\natexlab{b}}, \apjl, 705, L180

\bibitem[{{M{\"o}stl} {et~al.}(2009{\natexlab{c}}){M{\"o}stl}, {Farrugia},
  {Miklenic}, {Temmer}, {Galvin}, {Luhmann}, {Kilpua}, {Leitner},
  {Nieves-Chinchilla}, {Veronig}, \& {Biernat}}]{moe09}
{M{\"o}stl}, C., {et~al.} 2009{\natexlab{c}}, Journal of Geophysical Research
  (Space Physics), 114, 4102

\bibitem[{{M{\"o}stl} {et~al.}(2010){M{\"o}stl}, {Temmer}, {Rollett},
 {Farrugia}, {Liu}, {Veronig}, {Leitner}, {Galvin}, \& {Biernat}}]{moe10}
{M{\"o}stl}, C., {et~al.} 2010, \grl, 37, L24103

\bibitem[{{Nelder} \& {Mead}(1965)}]{nel65}
{Nelder}, J.~A., \& {Mead}, A. 1965, Computer Journal

\bibitem[{{Odstrcil}(2003)}]{ods03}
{Odstrcil}, D. 2003, Advances in Space Research, 32, 497

\bibitem[{{Patsourakos} \& {Vourlidas}(2009)}]{pat09}
{Patsourakos}, S., \& {Vourlidas}, A. 2009, \apjl, 700, L182

\bibitem[{{Rouillard} {et~al.}(2008){Rouillard}, {Davies}, {Forsyth}, {Rees},
  {Davis}, {Harrison}, {Lockwood}, {Bewsher}, {Crothers}, {Eyles}, {Hapgood},
  \& {Perry}}]{rou08}
{Rouillard}, A.~P., {et~al.} 2008, \grl, 35, 10110

\bibitem[{{Rouillard} {et~al.}(2009){Rouillard}, {Davies}, {Forsyth}, {Savani},
  {Sheeley}, {Thernisien}, {Zhang}, {Howard}, {Anderson}, {Carr}, {Tsang},
 {Lockwood}, {Davis}, {Harrison}, {Bewsher}, {Fr{\"a}nz}, {Crothers}, {Eyles},
  {Brown}, {Whittaker}, {Hapgood}, {Coates}, {Jones}, {Grande}, {Frahm}, \&
  {Winningham}}]{rou09}
{Rouillard}, A.~P., {et~al.}  2009, Journal of Geophysical Research (Space Physics), 114, 7106

\bibitem[{{Schmidt} \& {Bothmer}(1996)}]{sch96}
{Schmidt}, W.~K.~H., \& {Bothmer}, V. 1996, Advances in Space Research, 17, 369

\bibitem[{{Sheeley} {et~al.}(1999){Sheeley}, {Walters}, {Wang}, \&
  {Howard}}]{she99}
{Sheeley}, N.~R., {Walters}, J.~H., {Wang}, Y., \& {Howard}, R.~A. 1999, \jgr,
  104, 24739

\bibitem[{{Tappin} \& {Howard}(2009)}]{tap09}
{Tappin}, S.~J., \& {Howard}, T.~A. 2009, \ssr, 147, 55

\bibitem[{{Thernisien} {et~al.}(2009){Thernisien}, {Vourlidas}, \&
  {Howard}}]{the09}
{Thernisien}, A., {Vourlidas}, A., \& {Howard}, R.~A. 2009, \solphys, 256, 111

\bibitem[{{Tsurutani} {et~al.}(1988){Tsurutani}, {Smith}, {Gonzalez}, {Tang},
  \& {Akasofu}}]{tsu88}
{Tsurutani}, B.~T., {Smith}, E.~J., {Gonzalez}, W.~D., {Tang}, F., \&
  {Akasofu}, S.~I. 1988, \jgr, 93, 8519

\bibitem[{{Vourlidas} \& {Howard}(2006)}]{vou06}
{Vourlidas}, A., \& {Howard}, R.~A. 2006, \apj, 642, 1216

\bibitem[{{Williams} {et~al.}(2009){Williams}, {Davies}, {Milan}, {Rouillard},
  {Davis}, {Perry}, \& {Harrison}}]{wil09}
{Williams}, A.~O., {Davies}, J.~A., {Milan}, S.~E., {Rouillard}, A.~P.,
  {Davis}, C.~J., {Perry}, C.~H., \& {Harrison}, R.~A. 2009, Annales
  Geophysicae, 27, 4359

\bibitem[{{Wood} \& {Howard}(2009)}]{woo09}
{Wood}, B.~E., \& {Howard}, R.~A. 2009, \apj, 702, 901

\bibitem[{{Wood} {et~al.}(2010){Wood}, {Howard}, \& {Socker}}]{woo10}
{Wood}, B.~E., {Howard}, R.~A., \& {Socker}, D.~G. 2010, \apj, 715, 1524

\bibitem[{{Yamamoto} {et~al.}(2010){Yamamoto}, {Kataoka}, \& {Inoue}}]{yam10}
{Yamamoto}, T.~T., {Kataoka}, R., \& {Inoue}, S. 2010, \apj, 710, 456

\bibitem[{{Zhang} {et~al.}(2007){Zhang}, {Richardson}, {Webb}, {Gopalswamy},
  {Huttunen}, {Kasper}, {Nitta}, {Poomvises}, {Thompson}, {Wu}, {Yashiro}, \&
  {Zhukov}}]{zha07}
{Zhang}, J., {et~al.} 2007, Journal of Geophysical Research (Space Physics),
  112, 10102

\bibitem[{{Zhang} {et~al.}(2006){Zhang}, {Baumjohann}, {Delva}, {Auster},
  {Balogh}, {Russell}, {Barabash}, {Balikhin}, {Berghofer}, {Biernat},
  {Lammer}, {Lichtenegger}, {Magnes}, {Nakamura}, {Penz}, {Schwingenschuh},
  {V{\"o}r{\"o}s}, {Zambelli}, {Fornacon}, {Glassmeier}, {Richter}, {Carr},
  {Kudela}, {Shi}, {Zhao}, {Motschmann}, \& {Lebreton}}]{zha06}
{Zhang}, T.~L., {et~al.} 2006, \planss, 54, 1336

\end{thebibliography}

\clearpage




\clearpage

\clearpage






\end{document}